\documentclass[amsmath,amssymb,nofootinbib,notitlepage,superscriptaddress,twocolumn]{revtex4-2}

\usepackage{amsmath,graphicx,mathtools,setspace,subfigure}
\usepackage[margin=1in,letterpaper]{geometry}
\usepackage[colorlinks=true]{hyperref}

\usepackage[bb=dsserif]{mathalpha}
\usepackage{bm}

\newtagform{brackets}{[}{]}
\usepackage{amsfonts}
\usepackage{color}
\usepackage{graphicx}
\usepackage{amsmath}
\usepackage{amsthm}
\usepackage{mathrsfs}
\usepackage{bm}
\usepackage{comment}
\usepackage{braket}
\usepackage{ulem}
\usepackage[export]{adjustbox}

\newcommand{\brkt}[1]{\!\braket{#1}\!}

\newcommand{\pd}{\partial}

\begin{document}

\title{Decoherence from Horizons: General Formulation and Rotating Black Holes}
\author{Samuel E. Gralla}
\author{Hongji Wei}
\affiliation{Department of Physics, University of Arizona, Tucson, AZ 85721, USA}

\begin{abstract}
Recent work by Danielson, Satishchandran, and Wald (DSW) has shown that black holes---and, in fact, Killing horizons more generally---impart a fundamental rate of decoherence on all nearby quantum superpositions.  The effect can be understood from measurement and causality: An observer (Bob) in the black hole should be able to disturb outside quantum superpositions by measuring their superposed gravitational fields, but since his actions cannot (by causality) have this effect, the superpositions must automatically disturb themselves.  DSW calculated the rate of decoherence up to an unknown numerical factor for distant observers in Schwarzschild spacetime, Rindler observers in flat spacetime, and static observers in de Sitter spacetime.  Working in electromagnetic and Klein-Gordon analogs, we flesh out and generalize their calculation to derive a general formula for the precise decoherence rate for Killing observers near bifurcate Killing horizons.  We evaluate the rate in closed form for an observer at an arbitrary location on the symmetry axis of a Kerr black hole.  This fixes the numerical factor in the distant-observer Schwarzschild result, while allowing new exploration of near-horizon and/or near-extremal behavior.  In the electromagnetic case we find that the decoherence vanishes entirely in the extremal limit, due to the ``Black hole Meissner effect'' screening the Coulomb field from entering the black hole.  This supports the causality picture: Since Bob is unable to measure the field of the outside superposition, no decoherence is necessary---and indeed none occurs.
\end{abstract}

\maketitle

\section{Introduction}

Some of the greatest remaining conceptual challenges in theoretical physics involve the interplay among gravity, quantum mechanics, and measurement. 
 Thought experiments involving black holes can be particularly sharp, pushing cherished physical ideas to their limits.  In a recent paper  \cite{Danielson:2022tdw} remarkable for its simplicity and generality, Danielson, Satishchandran, and Wald (DSW) showed that the mere presence of a black hole induces a fundamental rate of decoherence on all nearby quantum superpositions.  In other words, it is not just difficult but in fact \textit{impossible} to avoid entanglement with degrees of freedom inside the black hole.

As surprising as this decoherence may seem, there is actually a very simple argument for why it should  occur.\footnote{This argument arises naturally out of prior thought experiments involving causality in flat spacetime \cite{Belenchia:2018szb,Danielson:2021egj} and indeed formed some of the original motivation for the DSW result (DSW, private communication).}  Suppose that Alice prepares a quantum superposition outside the black hole.  The superposed matter creates a superposed gravitational field, which penetrates into the black hole, where an observer, Bob, could choose to measure it.  By doing so, he has gained information about Alice's state and must therefore disturb it; however, by causality, his actions can have no effect whatsoever on her state.  Since a hypothetical Bob can in principle take an action that should disturb Alice's state, the inescapable conclusion is that her state must in fact \textit{disturb itself}: the DSW decoherence.

Though powerful, this simple argument still leaves one wanting for deeper understanding---some identifiable mechanism or at least a connection to more familiar physics.  DSW attribute the effect to ``emission of soft gravitons'' in light of a low-frequency approximation to the entangling graviton flux through the horizon.  But this is hardly emission in the conventional sense, given that the relevant quantum superpositions are held stationary as they steadily decohere.  A hint of deeper meaning might be found in the role of the surface gravity in the DSW calculation, suggesting that perhaps the decoherence can be understood in thermal terms.

Exploring and validating these interpretations---causality, soft radiation, thermal properties---will require detailed, general calculations of the decoherence.  Thus far, the effect has been considered for distant observers in the Schwarzschild spacetime \cite{Danielson:2022tdw}, Rindler observers in flat spacetime \cite{Danielson:2022sga}, and static observers in de Sitter spacetime \cite{Danielson:2022sga}.  These cases do not allow one to vary the surface gravity independently of other physical parameters and do not contain an extremal limit.  Furthermore, in all cases DSW only obtained the overall scaling of the rate, leaving a numerical factor undetermined.

In this paper we will derive a general formula for the precise decoherence rate for Killing observers near a bifurcate Killing horizon and apply it to an observer at an arbitrary location along the symmetry axis of a rotating black hole.  We  use the same basic setup as DSW, while fleshing out and/or generalizing some key elements of the calculation.  Our results confirm their findings (fixing unknown coefficients) and generalize the results to a context where the black hole mass, black hole temperature, and observer distance can be varied independently.  
However, our calculations are unfortunately limited to the electromagnetic (EM) and Klein-Gordon (KG) analogs.  We leave the more challenging case of gravitational decoherence to future work.\footnote{In the the gravitational case the stress-energy of Alice's laboratory must be taken into account.}

\begin{figure}
    \centering
    \includegraphics[width=\linewidth]{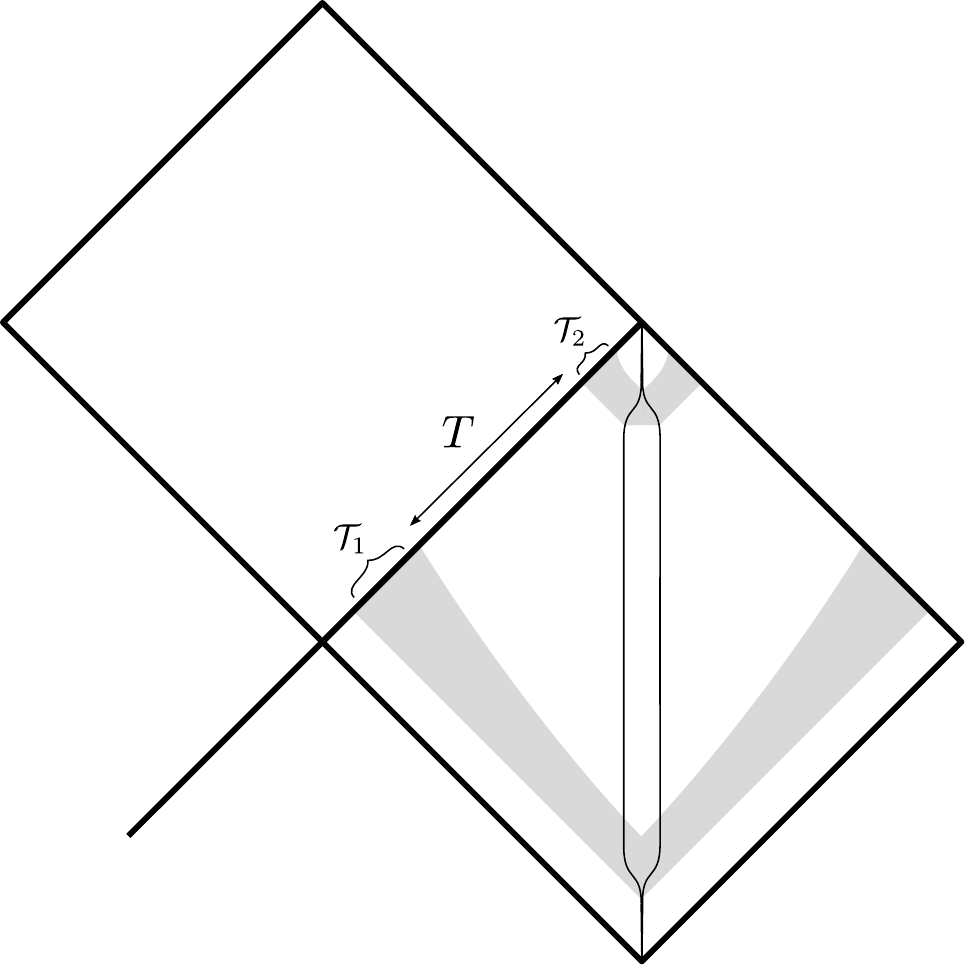}
    \caption{The DSW decoherence.  Alice creates a spatial superposition, holds it for Killing time $T$, and then closes it.  The superposed Coulomb fields are approximately stationary on the horizon for a time of order $T$, shown as a double-arrow.  At large $T$, the  coherence decreases exponentially like $e^{-\Gamma T}$, making complete decoherence unavoidable.  We find the precise rate to be $\Gamma=\frac{1}{2} C \kappa$, where $\kappa$ is the horizon surface gravity and $C$ is the ``decohering flux''  from the superposed Coulomb fields.}
    \label{fig:setup}
\end{figure}

We now describe the setup and the main results, focusing on the EM case.  As in DSW, we consider a localized (but macroscopic) charged object that is placed into in a spatial superposition, held for a (Killing) time $T$, and then recombined (Fig.~\ref{fig:setup}).  The charged object is treated semiclassically in that each branch of the superposition couples to the quantized EM field as a fixed classical source.  The sources are assumed to be on Killing orbits outside the horizon, and the photon field begins in the Unruh state associated with the Killing horizon.\footnote{By ``photon field'' we refer to the quantized retarded-minus-advanced field, which has no Coulomb degrees of freedom.  Here our approach differs technically from DSW, who instead subtract off Coulomb fields at initial and final times and consider the Hartle-Hawking vacuum, invoking low-frequency equivalence to Unruh.}

Each branch of the superposition has a different semi-classical source, so the two branches induce two different evolutions of the state $\ket{\psi}$ of the photon field, denoted $\ket{\psi_L}$ and $\ket{\psi_R}$ for ``left'' and ``right''.  If $\ket{\psi_L}$ and $\ket{\psi_R}$ become orthogonal after the experiment, then the experiment is maximally entangled with the photon field (a kind of ``environment'') and Alice's superposition has fully decohered.  Working in the limit that the superposition is maintained for a long time $T$, we calculate the final overlap of these states to be\footnote{Our calculation is non-extremal but we take the extremal limit using an estimate for the non-extremal rate under the condition $\kappa T \ll 1$, while still assuming that $T$ is much larger than all other timescales.}
\begin{align}\label{decoherence}
|\langle\psi_L|\psi_R\rangle| = \begin{cases}
\exp[{-\frac{C}{2} \kappa T}], & \kappa \neq 0 \\
(T/T_0)^{-C}, & \kappa = 0, \end{cases}
\end{align}
where $\kappa$ is the horizon surface gravity, $T_0$ is a non-universal timescale (dependent on the details of the transition), and $C$ is a constant we call the ``decohering flux''.

To present the formula for $C$, let $F_{\mu \nu}^R$ and $F_{\mu \nu}^L$ represent the stationary solutions associated with the right and left branches (respectively) of the superposition.  The difference between these fields contains ``which-path information'' about the superposition, and the important part turns out to be the ``radial fields'', i.e., the pullback of $F$ and its dual ${}^*F$ to a horizon cross-section $\mathcal{C}$.  We define
\begin{align}
    \widehat{\Delta E_r} & \equiv \frac{1}{2}\epsilon^{AB} ({}^*\!F^L_{AB}-{}^*\!F^R_{AB})|_{\mathcal{C}} \label{deltaEr} \\
    \widehat{\Delta B_r} & \equiv \frac{1}{2}\epsilon^{AB} (F^L_{AB}-F^R_{AB})|_\mathcal{C} \label{deltaBr},
\end{align}
where $x^A$ ($A=1,2$) are coordinates on $\mathcal{C}$, which has spatial metric $q^{AB}$, area form $\epsilon_{AB}$, area element $dS=\sqrt{q} d^2x^A$, derivative operator $\nabla_A$, and Laplacian $\nabla^2=q^{AB}\nabla_A \nabla_B$ with inverse  $\nabla^{-2}$.  In terms of these definitions, the decohering flux is given by\footnote{Eq.~\eqref{CEMformal} has been fully justified only for compact horizon cross-sections, but we expect it to hold in the non-compact case in spacetimes with suitable falloff properties.}

\begin{align}\label{CEMformal}
    C = \frac{1}{4\pi^2} \! \! \int \! \left[ (\nabla_A \nabla^{-2} \widehat{\Delta E_r})^2 + (\nabla_A \nabla^{-2} \widehat{\Delta B_r})^2 \right] \! dS,
\end{align}
where $(V_A)^2$ denotes $q^{AB} V_A V_B$.  
The KG result takes the same form \eqref{decoherence} with the simpler expression  $C=\pi^{-1}\int \widehat{\Delta\phi}^2 dS$ \eqref{C} for the decohering flux.

We are able to evaluate the decohering flux $C$ explicitly for a superposition on the symmetry axis of the Kerr spacetime.  In the EM case we consider a spatial superposition in the radial direction, while in the KG case we consider a superposition of different values of charge (which is possible because the KG charge can be time-dependent).  The KG and EM results are given in Eqs.~\eqref{CKerrKG} and Eq.~\eqref{CKerrEM} below, respectively, and are plotted in Fig.~\ref{fig:C}.

The extremal limit is particularly interesting.  From \eqref{decoherence} we see that if $\kappa \to 0$ with $C \neq 0$, the exponential decoherence goes over to a power-law behavior, whereas if both vanish then the decoherence vanishes entirely.  We see in Fig.~\ref{fig:C} that indeed both vanish ($C=\kappa=0$) for maximally spinning black holes ($a=M$) in the electromagnetic case, meaning the effect goes away completely.  This is a consequence of the ``black hole Meissner effect'' \cite{bivcak1985magnetic,Bini:2008iwy}  that maximally spinning black holes screen out any external fields.  The physical picture of ``Bob in a black hole'' thus holds up to detailed calculation: Since there is no Coulomb field for him to measure, there is no causal need for decoherence, and indeed the decoherence duly vanishes.

The remainder of the paper is organized as follows.  The first four sections treat the KG case in detail.  In Sec.~\ref{sec:QFT} we consider field quantization in the presence of a persistent semiclassical source and introduce the notion of particles as quanta of the retarded-minus-advanced field.  In Sec.~\ref{sec:experiment} we describe the superposition experiment and show that the decoherence is related to the ``expected number of entangling particles'', fleshing out the key ideas of DSW and extending them to the retarded-minus-advanced context.  In Sec.~\ref{sec:bifurcate} we express the decoherence as an integral on the past horizon.  This produces the main formal results, namely the decoherence rates and formulas for the decohering flux.  We calculate the decohering flux in the Kerr spacetime in Sec.~\ref{sec:Kerr}.  We consider the EM case in Sec.~\ref{sec:EM}, establishing some useful properties of a horizon-adapted gauge and repeating the steps of the KG calculation.  We use units with $G=c=\hbar = 1$ and $\epsilon_0=1/(4\pi)$, and our metric signature is $(-,+,+,+)$.

\begin{figure}
    \centering
    \includegraphics[width=\linewidth]{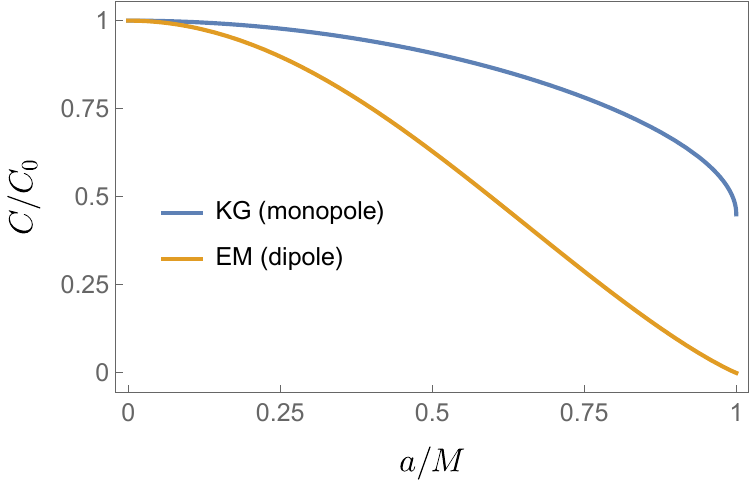}
    \caption{The decohering flux $C$ as a function of spin for an observer at $r_0=3M$ on the symmetry axis of a Kerr black hole, normalized by its value $C_0$ at $a=0$.  In the extremal limit, the EM flux vanishes as a result of the Black hole Meissner effect  (Fig.~\ref{fig:flux}).}
    \label{fig:C}
\end{figure}
\section{Klein-Gordon field with a classical source}\label{sec:QFT}

\begin{figure}
    \centering
    \includegraphics[width=.45\linewidth]{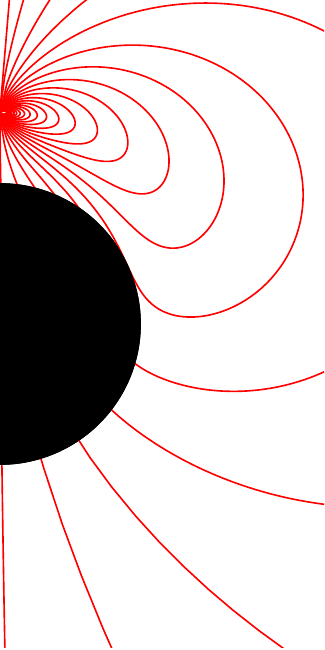}
    \ \ \ \ \ \  \includegraphics[width=.45\linewidth]{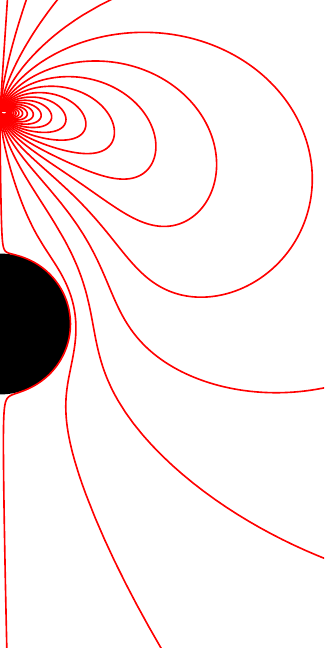}
    \caption{If a charge $q$ is placed in a spatial superposition of proper separation $d$, the ``which path information'' is contained in the difference of the Coulomb fields, which is field of a dipole $p=qd$.  Here we show the electric field lines (level sets of electric flux through a surface of revolution) for a dipole at $r_0=3M$ on the symmetry axis of a non-rotating (left) and maximally rotating (right) Kerr black hole.  (The black hole region $r<M+\sqrt{M^2-a^2}$ is shown in black.)  In the extremal case, the field does not penetrate inside (the black hole Meissner effect), and correspondingly there is no horizon-induced decoherence.
 }\label{fig:flux}
\end{figure}

In this section we review the Fock quantization of the Klein-Gordon field in the style of \cite{Birrell:1982ix}, first in the free case and then in the presence of a semiclassical source.  This establishes notation and physical interpretation for our analysis of the superposition experiment.  We work in a globally hyperbolic spacetime in coordinates $(t,x)$, where $t$ is constant on a family of Cauchy surfaces and $x$ refers to a collection of spatial coordinates.  We consider the case of three spatial coordinates, but the modifications for other dimensions are trivial.  

The free KG equation is
\begin{align}
    \Box \phi = 0,
\end{align}
where $\Box = g^{\mu \nu}\nabla_\mu \nabla_\nu$ and $\nabla$ is the metric-compatible covariant derivative.  Given two solutions $\phi_1$ and $\phi_2$, the KG product is defined as
\begin{align}\label{KGproduct}
    (\phi_1,\phi_2) & =  i \int_{\Sigma} \left(\ \! \overline{\phi}_1 \nabla_\mu \phi_2 - \phi_2\nabla_\mu \overline{\phi}_1  \right) n^\mu \sqrt{h} d^3x 
\end{align} 
where an overbar denotes complex conjugation and $n^\mu$ is the future-directed unit normal to a spacelike Cauchy surface with induced metric $h_{\mu \nu}$ and volume element $\sqrt{h} d^3x$.  The KG product is independent of the choice of surface, meaning in particular that its value is independent of $t$.  

We consider a complete set of mode solutions $\phi_i(t,x)$ normalized as
\begin{align}\label{orthogonal}
(\phi_i,\phi_j) = \delta_{ij}, \ \ (\overline{\phi}_i,\overline{\phi}_j) = -\delta_{ij},\ \ (\phi_i,\overline{\phi}_j) = 0. 
\end{align}
Here the index $i$ represents a collection of continuous or discrete indices, with $\delta_{ij}$ representing a product of Kronecker and/or Dirac deltas as appropriate.  If a solution is expanded as
\begin{align}
    \phi = \sum_i \left(c_i \phi_i + \overline{c}_i \overline{\phi}_i\right),
\end{align}
then its ``positive-frequency part'' is
\begin{align}\label{Kphi}
K \phi = \sum_i c_i \phi_i.
\end{align}
The map $K$ depends on the choice of mode functions. The KG product of a real solution (with itself) is always zero, but the KG product of its positive-frequency part is given by
\begin{align}\label{KphiKphi}
(K \phi, K \phi) = \sum_i |c_i|^2.
\end{align}

The quantum theory in the Heisenberg picture is obtained by promoting $\phi$ to a self-adjoint operator satisfying the field equations,
\begin{align}
    \Box \hat{\phi} = 0,
\end{align}
expressing the general solution in terms of a complete set of mode functions
\begin{align}\label{freefield}
\hat{\phi} = \sum_i \left( \hat{a}_i \phi_i + \hat{a}_i^\dagger \overline{\phi}_i \right),
\end{align}
imposing ladder-operator commutation relations on the coefficients $\hat{a}_i$,
\begin{align}\label{a commutators}   [\hat{a}_i,\hat{a}^\dagger_j] = \delta_{ij},\quad [\hat{a}_i,\hat{a}_j] = 0, \quad [\hat{a}^\dagger_i,\hat{a}^\dagger_j] = 0,
\end{align}
and building a Fock space in the usual way.  The product  $\hat{a}^\dagger_i \hat{a}_i$ is interpreted as the number operator for $\phi_i$-particles.  The notion of a particle thus depends on the choice of mode function.  This procedure is equivalent to canonical quantization \cite{Parker:2009uva}.

Now suppose that there is a fixed classical source $\rho(t,x)$, so that the Heisenberg-picture field now satisfies
\begin{align}\label{KGI}
    \Box \hat{\phi} = -4\pi \rho.
\end{align}
The form \eqref{freefield} is no longer a solution of the equations of motion; instead, we must add a particular solution to \eqref{KGI}.  This choice adds another freedom in the construction of the Fock space, beyond the choice of mode function.  The choice of mode function determines the notion of particle, while the two choices together---mode functions and particular solution---jointly determine the state of the field in the ``vacuum'' annihilated by $\hat{a}_i$.  For the semiclassical source to be the ``only influence'' on the state of the quantum field, the natural condition is the absence of incoming radiation, corresponding to the (classical) retarded solution $\phi^{\rm ret}$,\footnote{By assuming the existence of the retarded solution we restrict to situations where the source is suitably regular at early times.}
\begin{align}\label{phi}
    \hat{\phi} & = \sum_i \left( \hat{a}_i \phi_i + \hat{a}_i{}^\dagger \overline{\phi}_i \right) + \phi^{\rm ret} \hat{\mathbb{1}}.
\end{align}
where $\hat{\mathbb{1}}$ is the identity. We again impose the commutation relations \eqref{a commutators}.  The operators $\hat{a}_i$ define an ``in'' Fock space $\mathcal{F}_{\rm in}$ with an associated ``in'' vacuum,
\begin{align}\label{invac}
\hat{a}_i \ket{\rm in} = 0,
\end{align}
which has no in-particles.  However, this notion of particle is unnatural at late times, since a field configuration with outgoing radiation would be deemed to have no particles.  Instead, at late times the natural expression of $\hat{\phi}$ is
\begin{align}\label{field}
    \hat{\phi} & = \sum_i \left( \hat{b}_i \phi_i + \hat{b}_i{}^\dagger \overline{\phi}_i \right) + \phi^{\rm adv} \hat{\mathbb{1}},
\end{align}
where the use of the \textit{advanced} field $\phi^{\rm adv}$ guarantees that the $b$-vacuum has no \textit{outgoing} radiation, as desired if one wishes to have a quantized description of outgoing radiation.  These operators define an ``out'' Fock space $\mathcal{F}_{\rm out}$ with an associated ``out'' vacuum,
\begin{align}\label{outvac}
\hat{b}_i \ket{\rm out} = 0,
\end{align}
which has no out-particles.  Notice that we use the same mode functions $\phi_i$ at early and late times.  This in effect neglects any particle creation due to the spacetime itself.  Such effects can easily be added on with a Bogoliubov transformation.

In general the source $\rho$ will create particles in the sense that $\bra{\rm in}\hat{N}_{\rm out}\ket{\rm in} \neq 0$, where $\hat{N}_{\rm out}=\hat{b}_i^\dagger \hat{b}_i$. We can express the particle creation by noticing that
\begin{align}\label{bcalpha}
\hat{a}_i = \hat{b}_i + \alpha_i\hat{\mathbb{1}},
\end{align}
where $\alpha_i$ are the coefficients of the retarded-minus-advanced solution,
\begin{align}\label{phiRA}
    \phi^{\rm ret} - \phi^{\rm adv} = \sum_i\left( \alpha_i \phi_i + \overline{\alpha}_i \overline{\phi}_i \right).
\end{align}
Since the retarded-minus-advanced solution is a homogeneous solution to the Klein-Gordon equation, these coefficients $\alpha_i$ are constants independent of time (and space).  It then follows that the expected number of particles created in each mode $i$ is given by 
\begin{align}\label{particles}
    \bra{\rm in}\hat{b}_i^\dagger \hat{b}_i\ket{\rm in} = |\alpha_i|^2.
\end{align}
In light of Eqs.~\eqref{Kphi} and \eqref{KphiKphi}, we see that the \textit{total} expected particle number is given by
\begin{align}\label{particlesKG}
\sum_i |\alpha_i|^2 = (K(\phi^{\rm ret}-\phi^{\rm adv}),K(\phi^{\rm ret}-\phi^{\rm adv})).
\end{align}
That is, the expected number of particles is the KG product of the positive-frequency part of the retarded-minus-advanced solution.

In the next section we will consider the response of the KG field to a quantum superposition of semiclassical sources that agree at early times.  For this purpose it is useful introduce the Schr\"odinger picture defined relative to some early time $t_0$.  We denote the Schr\"odinger-picture operator by $\hat{\phi}_0$ and the initial Schr\"odinger-picture state by $\ket{\psi_0}$,
\begin{align}
    \hat{\phi}_0 & = \hat{\phi}(t_0,x) \label{phi0} \\
    \ket{\psi_0} & = \ket{\rm in}. \label{psi0}
\end{align}
The Schr\"odinger-picture state evolves as
\begin{align}
    \ket{\psi} = \hat{U}(t,t_0) \ket{\psi_0},
\end{align}
where the quantum-mechanical propagator $\hat{U}$ satisfies 
\begin{align}
    \hat{\phi}_0 = \hat{U}(t,t_0) \hat{\phi}(t,x) \hat{U}(t,t_0)^\dagger, \label{phi0U}
\end{align}
with $\hat{\phi}$ given in Eq.~\eqref{phi}.

\section{The superposition experiment}\label{sec:experiment}

Let $\mathcal{H}_{\rm matter}$ denote the Hilbert space of the matter degrees of freedom manipulated by our intrepid experimentalist.  She prepares a superposition state 
\begin{align}
\ket{\Psi} = \frac{1}{\sqrt{2}} \left( \ket{\Psi_L} + \ket{\Psi_R} \right),
\end{align}
with the following properties.  First, the constituent states are perfectly distinguishable, 
\begin{align}
    \braket{\Psi_L|\Psi_R} = 0.
\end{align}
Second, their charge densities can be treated semiclassically, 
\begin{align}
    \hat{\rho} \ket{\Psi_I} \approx \rho_I(t,x) \ket{\Psi_I} \qquad (I=L,R).
\end{align}
Here $\hat{\rho}$ is the charge operator and $\rho_I$ is its expectation value in the (normalized) state $\ket{\Psi_I}$.  Third, the two semiclassical charge densities are identical at sufficiently early and late times,
\begin{align}
   \rho_L = \rho_R \qquad \textrm{(early and late times).}
\end{align}

We name the states left and right since we imagine a spatial superposition, following DSW.  Such a superposition can be achieved if matter with an embedded spin is sent through a Stern-Gerlach apparatus, held in spatial superposition for a time $T$, and then sent through a reversing Stern-Gerlach apparatus \cite{Belenchia:2018szb}.  However, in the KG case the charge of the particle is not conserved, and we will actually consider the simpler situation where the ``left'' and ``right'' states are two different time-evolutions of the charge, without any spatial separation.  These distinctions are unimportant until Sec.~\ref{sec:Kerr} below.

Now consider the evolution of the system including the Klein-Gordon coupling.  By the assumption of no backreaction, each state $\ket{\Psi_I(t)}$ in the superposition induces a corresponding evolution $\ket{\psi_I(t)}$ in the field via the corresponding semiclassical source $\rho_I$.  Because the sources agree at early times, the corresponding retarded solutions $\phi^{\rm ret}_I$ agree at the initial time $t_0$.  Using a fixed set of mode functions $\phi_i$, the Shr\"odinger-picture field $\hat{\phi}$ [see \eqref{phi0} and \eqref{phi}] is therefore independent of the choice of right or left and can be identified with the single Shr\"odinger-picture field operator in the superposition experiment.  Similarly, the time-independent annihilation operators $\hat{a}_i$ define the shared initial state of the field via
\begin{align}\label{state}
    \hat{a}_i\ket{\psi_0} = 0.
\end{align}

This construction results in an initial tensor product state for the joint evolution,
\begin{align}
    \ket{\Upsilon} = \frac{1}{\sqrt{2}} \left( \ket{\Psi_L} + \ket{\Psi_R} \right) \otimes \ket{\psi_0}, \quad t=t_0.
\end{align}
The interpretation is that Alice has carefully isolated her apparatus.  As she conducts the superposition experiment, the photon field reacts differently do the right and left branches, and the system as a whole evolves as
 \begin{align}
    \ket{\Upsilon} = \frac{1}{\sqrt{2}} \left( \ket{\Psi_L} \otimes \ket{\psi_L} + \ket{\Psi_R} \otimes \ket{\psi_R}\right).
\end{align}
If the left and right field states become orthogonal, then the superposition has decohered.

To express the evolution of the field states we add an index to the propagator $\hat{U}$,
\begin{align}\label{psiI}
    \ket{\psi_I} = \hat{U}_I(t,t_0) \ket{\psi_0}.
\end{align}
Whereas in Sec.~\ref{sec:QFT} $\hat{U}$ was \textit{the} propagator for the problem with a fixed semiclassical source, here $\hat{U}_I$ is the evolution operator for a particular part of a particular state.  More precisely, if the matter states $\ket{\Psi}$ evolve by $\hat{\mathcal{U}}$, then $\ket{\Psi_I} \otimes \ket{\psi_I}$ evolves by $\hat{\mathcal{U}}\otimes\hat{U}_I$. 

The Heisenberg-picture operators of Sec.~\ref{sec:QFT} depend on the source and are not associated with any picture (Heisenberg or otherwise) in the superposition problem.  Instead, these are viewed as useful definitions to obtain information about $\hat{U}_I$.  From Eq.~\eqref{phi}, we have
\begin{align}\label{phiI}
    \hat{\phi}_I & = \sum_i \left( \hat{a}_i \phi_i + \hat{a}_i{}^\dagger \overline{\phi}_i \right) + \phi^{\rm ret}_I\hat{\mathbb{1}},
\end{align}
and from \eqref{phi0} we have
\begin{align}\label{phi0LR}
    \hat{\phi}_0(x) = \hat{U}_I(t,t_0) \hat{\phi}_I(t,x) \hat{U}_I(t,t_0)^\dagger.
\end{align}

Our goal is to calculate
\begin{align}\label{niceoverlap}
\lim_{t \to \infty}|\langle \psi_L|\psi_R \rangle| = |\langle \psi_0| \hat{D} |\psi_0 \rangle|,
\end{align}
where in light of Eq.~\eqref{psiI}, we define 
\begin{align}
\hat{D} = \lim_{t \to \infty} \hat{U}_L^\dagger(t,t_0) \hat{U}_R(t,t_0).
\end{align}
To find this operator we note that since the LHS of \eqref{phi0LR} is independent of $I$, we have
\begin{align}
    \hat{U}_L \hat{\phi}_L \hat{U}_L^\dagger = \hat{U}_R \hat{\phi}_R \hat{U}_R^\dagger,
\end{align}
implying
\begin{align}\label{RL}
    \hat{\phi}_R = \hat{U}_R^\dagger \hat{U}_L \hat{\phi}_L \hat{U}_L^\dagger \hat{U}_R,
\end{align}
where we have suppressed the $(t,x)$ dependence of the fields and the $(t,t_0)$ dependence of the propagators.  From \eqref{phiI} we have
\begin{align}
    \hat{\phi}_R = \hat{\phi}_L + \left(\phi_R^{\rm ret} - \phi_L^{\rm ret}\right)\hat{\mathbb{1}},
\end{align}
which combined with \eqref{RL} gives
\begin{align}
    \hat{U}_R^\dagger \hat{U}_L \hat{\phi}_L \hat{U}_L^\dagger \hat{U}_R = \hat{\phi}_L + \left(\phi_R^{\rm ret}-\phi_L^{\rm ret}\right)\hat{\mathbb{1}}.
\end{align}
At late times $t \to \infty$, the LHS is $\hat{D}^\dagger \hat{\phi}_L \hat{D}$.  Since the advanced fields agree at late times, we may add $\phi^{\rm adv}_L-\phi^{\rm adv}_R$ to the RHS in this limit.  We thus have
\begin{align}\label{alexu}
    \hat{D}^\dagger \hat{\phi}_L \hat{D} & = \hat{\phi}_L + \Delta \phi \hat{\mathbb{1}}, 
\end{align}
where we introduce
\begin{align}\label{Deltaphi}
    \Delta \phi & \equiv (\phi_R^{\rm ret} - \phi_R^{\rm adv}) - ( \phi_L^{\rm ret} - \phi_L^{\rm adv}).
\end{align}
Eq.~\eqref{alexu} holds at all times, although $\hat{D}$ was defined using a late-time limit.  At late times, Eq.~\eqref{alexu} reduces to $\hat{D}^\dagger \hat{\phi}_L \hat{D}=\hat{\phi}_R$, showing that $\hat{D}$ displaces the late-time left field to the late-time right field.

The form \eqref{alexu} is convenient because the presence of the retarded-minus-advanced fields makes $\Delta \phi$ a source-free solution for all times.  Its mode coefficients are therefore time-independent, and we can think of $\hat{D}$ as a time-independent displacement operator.  According to the general theory of coherent states \cite{Glauber:1963tx}, every operator that obeys \eqref{alexu} may be represented as 
\begin{align}\label{Dcong}
    \hat{D} \cong \exp \left[ \sum_i \hat{a}_i^\dagger (\alpha_i^R - \alpha_i^L) - \hat{a}_i (\overline{\alpha}_i^R - \overline{\alpha}_i^L)\right],
\end{align}
where $\cong$ indicates equality up to phase and $\alpha^I_i$ are the time-independent mode coefficients of $(\phi_I^{\rm ret} - \phi_I^{\rm adv})$.  Using \eqref{state}, one can then calculate the overlap \eqref{niceoverlap} as
\begin{align}\label{niceRL}
 |\langle \psi_0| \hat{D} |\psi_0 \rangle| = \exp\left(-\frac{1}{2}\sum_i|\alpha_i^R-\alpha_i^L|^2\right).
\end{align}

Recall from \eqref{particles} that $\sum_i|\alpha_i|^2$ can be interpreted as the expected number of particles produced by the semiclassical source.  Here it is the left-right \textit{difference} $\alpha_i^R-\alpha_i^L$ that appears, which by linearity can be interpreted as the expected number of particles produced by the \textit{difference} of the left and right semiclassical sources.  Since these particles evidently determine the extent to which the system decoheres, it is natural to call them \textit{entangling particles} following DSW.  We therefore define $\langle N_e \rangle$, the ``expected number of entangling particles'', as
\begin{align}\label{Ne}
    \langle N_e \rangle & = \sum_i|\alpha_i^R-\alpha_i^L|^2 ,
\end{align}
which is also the KG product of the positive-frequency part of the left-right difference of the retarded minus advanced solutions \eqref{Deltaphi},
\begin{align}\label{Ne2}
    \langle N_e \rangle = (K\Delta \phi,K\Delta \phi).
\end{align}
Putting everything together, Eqs.~\eqref{niceoverlap}, \eqref{niceRL}, and \eqref{Ne} become
\begin{align}\label{largetdecoherence}
    \lim_{t\to \infty}|\!\braket{\psi_L|\psi_R}\!| = e^{-\langle N_e\rangle/2}
\end{align}
This reproduces the physical picture of Ref.~\cite{Belenchia:2018szb} and DSW: if the experiment emits entangling particles, the superposition decoheres.  Notice that near-complete decoherence occurs with the emission of just a handful of particles.

In computing $\langle N_e \rangle$ using \eqref{Ne2}, the KG product may be computed on any Cauchy slice, since $\Delta \phi$ is a homogeneous solution.  However, if we use a slice at late times, where the advanced solutions agree ($\phi^{\rm adv}_L=\phi^{\rm adv}_R$), then only the retarded solutions appear.  That is, the expected number of entangling particles is the KG product of the positive-frequency part of the difference between the left and right retarded solutions, evaluated at late times.  This is the form of the result obtained in DSW.

In calculations we will find it convenient to instead push the Cauchy surface to early times, where (for the Unruh state) the non-vanishing mode functions are purely positive frequency.  In this case it is the retarded solutions that agree ($\phi^{\rm ret}_L=\phi^{\rm ret}_R$), and so only the advanced solutions appear.  That is, we may also compute $\langle N_e \rangle$ as the KG product of the positive-frequency part of the difference between the left and right advanced solutions, evaluated at early times.

\section{Entangling particles on a bifurcate Killing horizon in the Unruh state}\label{sec:bifurcate}

\begin{figure}
    \includegraphics[width=.8\linewidth]{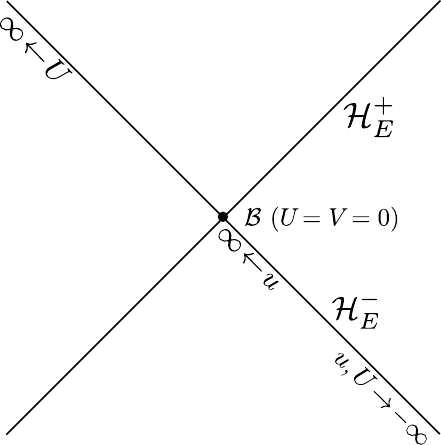}
    \centering
    \caption{Bifurcate Killing horizon}\label{fig:bifurcate}
\end{figure}

We now restrict to spacetimes with a bifurcate Killing horizon (Ref.~\cite{Kay:1988mu} and App.~\ref{sec:killing}), in which a future horizon $\mathcal{H}^+$ and a past horizon $\mathcal{H}^-$ intersect at a bifurcation surface $\mathcal{B}$.  We label the horizon generators of $\mathcal{H}^\pm$ by their coordinates $x^A$ on $\mathcal{B}$ and use an affine parameter $U\in (-\infty,\infty)$ on the past horizon and $V\in (-\infty,\infty)$ on the future horizon, such that $U=V=0$ is $\mathcal{B}$.  Thus $\mathcal{H}^+$ is described by coordinates $(V,x^A)$ and $\mathcal{H}^-$ is described by coordinates $(U,x^A)$.  In black hole spacetimes with the conventional definitions of $U$ and $V$, the regions of $\mathcal{H}_\pm$ bordering the exterior are $U<0$ on $\mathcal{H}^-$ and $V>0$ on $\mathcal{H}_+$.  We will denote these portions as $\mathcal{H}_{\pm}^E$.  We denote the horizon surface gravity by $\kappa$ and introduce Killing time coordinates $u$ and $v$ by
\begin{align}\label{UV}
    U = - e^{-\kappa u}, \qquad V = e^{\kappa v},
\end{align}
In particular, $\mathcal{H}^-_E$ is covered by $-\infty < u <\infty$, while $\mathcal{H}_+^E$ is covered by $-\infty < v <\infty$.  These coordinates $u$ and $v$ extend into the conventional exterior region of a black hole spacetime.  Some of these properties are illustrated in Fig.~\ref{fig:bifurcate}.

\subsection{KG product in terms of affine-time Fourier transform}\label{sec:KG product in affine time FT}

The formula \eqref{KGproduct} for the KG product holds on a spacelike surface.  We will be interested in the limit where a portion of this surface approaches the past horizon $\mathcal{H}^-$.  One can check that the limiting product is 
\begin{align}\label{KGonNull}
(\phi_1,\phi_2)_{\mathcal{H}^-} & = i \int \ell^\mu \left(\ \! \overline{\phi}_1 \nabla_{\mu} \phi_2 -  \phi_2\nabla_\mu \overline{\phi}_1\right) dS d\lambda,
\end{align}
where $dS$ is the volume element on the bifurcation surface $\mathcal{B}$, while $\ell^\mu=d x^\mu / d\lambda$ is tangent to the horizon generators $x^\mu(\lambda)$ with $\lambda$ some parameter that is constant on $\mathcal{B}$.  Choosing $\lambda=U$ gives
\begin{align}\label{KGonH}
    (\phi_1,\phi_2)_{\mathcal{H}^-} & = i \int_{\mathcal{H}^-} \!\!\! \left( \ \!\overline{\phi}_1 \pd_{U} \phi_2 - \phi_2\pd_U \overline{\phi}_1 \right) dS \ \!dU. 
\end{align}

Consider now a set of modes $\phi_i$ used for quantization.  Suppose in particular that the subset of such modes that are non-vanishing on the past horizon are purely positive-frequency with respect to affine time $U$ when evaluated there.  (This situation arises for black holes formed from gravitational collapse; we will describe it as working in the ``Unruh state'' \cite{Unruh:1976db}).  Describing each non-vanishing mode by an affine frequency $\Omega>0$ and a set of (continuous or discrete) spatial indices $L$, we have
\begin{align}\label{unruhmode}
    \phi_{\Omega L}|_{\mathcal{H}^-} = \frac{e^{-i \Omega U}}{\sqrt{4\pi\Omega}} Y_{L}(x^A),
\end{align}
where $Y_{L}$ is some complete set of modes for functions $f(x^A)$ which are orthonormal with respect to $dS$,\footnote{In practice, the most useful set of spatial mode functions may depend on $\Omega$, i.e., may be some set $Y_{\Omega L}(x^A)$.  The details of the mode functions play no role in our considerations, so for simplicity we assume they depend only on $L$.} 
\begin{align}
    \int \overline{Y}_L(x^A) Y_{L'}(x^A) dS = \delta_{L L'}.
\end{align}
The prefactor $(4\pi\Omega)^{-1/2}$ in \eqref{unruhmode} guarantees that the modes are properly normalized in the sense of \eqref{orthogonal}.  The mode coefficient $c_{\Omega L}$ is then evaluated to be 
\begin{align}
c_{\Omega L} & = (\phi_{\Omega L},\phi)|_{\mathcal{H}^-} \nonumber \\ 
& = \sqrt{\frac{\Omega}{\pi}} \int dS \int_{-\infty}^\infty dU \phi|_{\mathcal{H}^-} e^{-i \Omega U} Y_L(x^A).
\end{align}
Noting that $c_{\Omega L}$ is proportional to the affine-time Fourier transform of each spatial mode of the field, we can express the Fourier transform as a sum over modes,
\begin{align}
    \tilde{\Phi}(\Omega,x^A) = \sum_L \sqrt{\frac{\pi}{\Omega}}c_{\Omega L} \overline{Y}_L(x^A),
\end{align}
where our Fourier conventions are 
\begin{align}\label{Phitilde}
    \tilde{\Phi}(\Omega,x^A) = \int_{-\infty}^{\infty} \phi|_{\mathcal{H}^-}(U,x^A) e^{- i \Omega U} dU.
\end{align}
Integrating over the spatial directions, we have
\begin{align}
    \int |\tilde{\Phi}(\Omega,x^A)|^2 dS = \sum_L \frac{\pi|c_{\Omega L}|^2}{\Omega},
\end{align}
using the orthonormality of the harmonics.  The KG product of the positive-frequency part of a field $\phi$ is thus
\begin{align}
    (K\phi,K\phi)_{\mathcal{H}^-} & =  \int_0^{\infty} d \Omega \sum_L |c_{\Omega L}|^2 \\
    & = 2 \int dS \int_0^\infty \frac{\Omega d\Omega}{2\pi} |\tilde{\Phi}(\Omega,x^A)|^2.\label{doubleKphiKphi}
\end{align}
The range of the $\Omega$-integral follows from the fact that the mode functions are purely positive frequency (defined only for $\Omega>0$).  

The future-horizon analog of Eq.~\eqref{doubleKphiKphi} was used by DSW in the electromagnetic and gravitational cases in the Schwarzschild spacetime \cite{Danielson:2022tdw}.  The future-horizon version holds for the Hartle-Hawking state, but DSW argue that the associated decoherence results also apply to the Unruh state, citing low-frequency equivalence between the two.

We will find it helpful to introduce another form of Eq.~\eqref{doubleKphiKphi}.  Using the definition \eqref{Phitilde} of the Fourier transform, we can introduce two integrals over affine time,
\begin{align}
    & (K\phi, K\phi)_{\mathcal{H}^-}  = \frac{1}{\pi} \int dS \int_{-\infty}^\infty d U_1 \int_{-\infty}^\infty dU_2 \nonumber \\ & \ \phi(U_1,x^A) \phi(U_2,x^A) \int_0^\infty \Omega  e^{-i\Omega(U_1-U_2)} d\Omega.
\end{align}
This last integral can be performed using the identity
\begin{align}
\int_{-\infty}^\infty \frac{e^{i \omega x}}{(x-i \epsilon)^2} dx = - 2 \pi \omega H(\omega),
\end{align}
which is easily checked by contour integration.  Here $H(\omega)$ is the Heaviside function, equal to $1$ for $\omega>0$ and $0$ for $\omega<0$. Inverting this Fourier transform, we find
\begin{align}
\frac{1}{(x-i \epsilon)^2}= - \int_{-\infty}^\infty \omega d\omega H(\omega) e^{-i\omega x} ,
\end{align}
which is the needed integral.  Eq.~\eqref{doubleKphiKphi} then becomes
\begin{align}\label{whitey}
    (K\phi, K\phi)_{\mathcal{H}^-} \! = \! \frac{-1}{\pi} \! \! \int \! \! dS d U_1 dU_2 \frac{\phi(U_1,x^A) \phi(U_2,x^A)}{(U_1-U_2 - i \epsilon)^2}.
\end{align}
This formula also arises in the rigorous approach to QFT used in \cite{Kay:1988mu} -- see equation 4.13 therein.

\subsection{KG product in terms of Killing-time Fourier transform}\label{sec:KG product killing time FT}

Up to now we have considered the entirety of $\mathcal{H}^-$, including both the ``exterior'' $U<0$ and ``interior'' $U>0$.   However, our application of the formula involves fields that are non-zero only in the exterior part $U<0$, denoted $\mathcal{H}^-_E$.  For such fields we may reduce the integration range of $U_1$ and $U_2$ to negative values only, and we may change variables to the Killing time $u$ related by $U=-e^{-\kappa u}$ \eqref{UV}.  After some algebra we obtain
\begin{align}
    (K\phi, & K\phi)_{\mathcal{H}^-_E} =  \frac{-\kappa^2}{4\pi} \int dS \int_{-\infty}^\infty du_1 \int_{-\infty}^\infty du_2 \nonumber \\ 
    & \frac{\phi(u_1,x^A) \phi(u_2,x^A)}{\sinh^2\!\left( \tfrac{1}{2} \kappa(u_1-u_2)-i\epsilon\right)},\label{whitey2}
\end{align}
where $\phi$ is evaluated on $\mathcal{H}^-_E$ and regarded as a function of $(u,x^A)$.  We write $\mathcal{H}^-_E$ on the LHS to remind the reader that this integral ranges over only the exterior portion of the past horizon.  However, the full horizon is still in some sense involved through the $K$ map, which refers to positive frequency with respect to $U \in (-\infty,\infty)$.

Next we reintroduce the Fourier transform, this time with respect to Killing time,
\begin{align}\label{phitilde}
    \tilde{\phi}(\omega,x^A) = \int_{-\infty}^\infty \phi|_{\mathcal{H}^-_E}(u,x^A) e^{-i \omega u} du.
\end{align}
We may now invert for $\phi|_{\mathcal{H}^-_E}(u,x^A)$ and plug in to both instances of $\phi$ in Eq.~\eqref{whitey2}.  After changing variables to $u=u_1$ and $\Delta =u_2-u_1$, we see that the $u$ integral reduces to a delta function in frequency, leaving
\begin{align}
    (K\phi,  K\phi)_{\mathcal{H}^-_E} = & \frac{-\kappa^2}{8\pi^2} \int dS \int_{-\infty}^\infty d\omega |\tilde{\phi}(\omega,x^A)|^2 \nonumber \\ &  \int_{-\infty}^\infty d\Delta \frac{e^{-i\omega \Delta}}{\sinh^2\!\left( \tfrac{1}{2} \kappa \Delta -i\epsilon\right)}. 
\end{align}
The last integral may be performed by contour integration, giving 
\begin{align}\label{best}
    (K\phi, & K\phi)_{\mathcal{H}^-_E} \\ = & \frac{1}{\pi} \int dS \int_{-\infty}^\infty d\omega |\tilde{\phi}(\omega,x^A)|^2 \frac{\omega}{\exp\left(2\pi \omega/\kappa\right)-1}, \nonumber
\end{align}
which may be compared with \eqref{doubleKphiKphi}.  

From Eqs.~\eqref{Ne2}, \eqref{best}, and \eqref{phitilde} and \eqref{Deltaphi} the contribution from $\mathcal{H}^-_E$ to the expected number of entangling photons is given by 
\begin{align}\label{NeKilling}
    \langle N_e \rangle_{\mathcal{H}^-_E} = \frac{1}{\pi} \int dS \int_{0}^\infty d\omega |\Delta \tilde{\phi}|^2 \omega \coth\left( \frac{\pi \omega}{\kappa} \right),
\end{align}
where $\Delta \tilde{\phi}$ is the Killing-time Fourier transform of $\Delta \phi$ \eqref{Deltaphi} evaluated on the horizon, 
\begin{align}\label{Deltaphitilde}
    \Delta\tilde{\phi} = \int_{-\infty}^\infty \Delta \phi|_{\mathcal{H}^-_E} e^{-i \omega u} du.
\end{align}
The left and right retarded solutions agree at early times, so they do not contribute to $\Delta \phi$ when evaluated on $\mathcal{H}^-_E$,
\begin{align}
\Delta \phi|_{\mathcal{H}^-_E} = -(\phi^{\rm adv}_R-\phi^{\rm adv}_L)|_{\mathcal{H}^-_E}.\label{annoyingsign}
\end{align}

Eq.~\eqref{NeKilling} has a thermal interpretation.  If the calculation were repeated using Boulware modes (positive frequency with respect to $u$ on $\mathcal{H}^-_E$) to quantize the field in the exterior, then Eq.~\eqref{NeKilling} would appear without the $\coth$ factor.  The difference between the Unruh case under consideration and the Boulware case \cite{frolov_renormalized_1989,boulware_quantum_1975,Unruh_kerr_1974,zeldovich_1971} is thus an integral multiplied by $\coth(\pi\omega/\kappa)-1$, which is a thermal factor $2/(\exp[\omega/T_H]-1)$ where $T_H = \kappa/(2\pi)$ is the horizon temperature.

\subsection{Estimate of entangling particle number}\label{sec:killing-time-estimate}

For the superposition experiment in a spacetime with sufficient decay properties, the left and right fields will agree at early and late times, meaning that $\Delta \phi$ will vanish at early and late times.  If the superposition is maintained for a sufficiently long time, then the left and right fields will be roughly constant over most of the time the superposition is maintained, and $\Delta \phi$ is similarly constant over a corresponding  lapse of time.  We will denote this constant by $\widehat{\Delta \phi}(x^A)$.  Again assuming the relevant decay properties, it can be computed from solutions where the sources persist forever,
\begin{align}\label{widephidef}
    \widehat{\Delta \phi}(x^A) = - \left.\left( \phi_R^{\rm stat} - \phi^{\rm stat}_L \right)\right|_{\mathcal{C}},
\end{align}
where $\phi_R^{\rm stat}$ and $\phi_L^{\rm stat}$ are stationary (invariant under the horizon-generating Killing field\footnote{If the source were instead invariant under another Killing field, the KG field would not in general be constant on each horizon generator $x^A$ during the superposition time.}) solutions with stationary sources $\rho_R^{\rm stat}$ and $\rho_L^{\rm stat}$ corresponding to the superposition.  We write evaluation on a horizon cross-section $\mathcal{C}$ (as opposed to the past horizon $\mathcal{H}^-_E$) to emphasize that these are stationary solutions.  We include a minus sign in the definition \eqref{widephidef} to match the minus sign in \eqref{annoyingsign}, which originates from the use of retarded \textit{minus} advanced fields in the difference field \eqref{Deltaphi}.  This is purely a cosmetic issue since only the square of $\widehat{\Delta \phi}$ will appear in the decohering flux, Eq.~\eqref{C} below.

We may thus think of $\Delta \phi$ as transitioning from zero to $\widehat{\Delta \phi}$ then back to zero.  We will work with Killing time $u$ and denote the Killing timescales as follows:
\begin{itemize}
\setlength\itemsep{-.5em}
\item[] $T_b$: The largest ``background'' timescale (associated with propagation on the background spacetime) \\
\item[] $T_t$: The ``transition'' timescale (associated with opening or closing the superposition) \\
\item[] $T$: The timescale for which the superposition is maintained
\end{itemize}
Finally we define 
\begin{align}
    \mathcal{T} = \textrm{max}(T_t,T_b).
\end{align}
which represents the timescale for the field on the horizon to transition.  

Our goal is to estimate the integral \eqref{NeKilling} in the regime
\begin{align}
    T \gg \mathcal{T}.
\end{align}
To do so we introduce an intermediate frequency $\omega_c$ satisfying
\begin{align}\label{scales}
    \frac{1}{T} \ll \omega_c \ll \frac{1}{\mathcal{T}}. 
\end{align}
For example, if $1/T \sim \epsilon$ then we can take $\omega_c \sim \epsilon^{1/2}$ as $\epsilon \to 0$.  We split the integral \eqref{NeKilling} into two pieces,
\begin{align}\label{split}
    \langle N_e \rangle_{\mathcal{H}^-_E} = I_1(T) + I_2(T),
\end{align}
with
\begin{align}
    I_1(T) & = \frac{1}{\pi} \int dS \int_{0}^{\omega_c} d\omega |\Delta \tilde{\phi}|^2 \omega \coth\left( \frac{\pi \omega}{\kappa} \right) \label{F1} \\
    I_2(T) & = \frac{1}{\pi} \int dS \int_{\omega_c}^\infty d\omega |\Delta \tilde{\phi}|^2 \omega \coth\left( \frac{\pi \omega}{\kappa} \right).\label{F2}
\end{align}

In the first integral \eqref{F1} we have $\omega \mathcal{T} \ll 1$ on the entire range of integration, so that variations in $\Delta \phi$ on scales $\lesssim \mathcal{T}$ are not visible in the Fourier transform.  In this regime we may approximate $\Delta \tilde{\phi}$ as the Fourier transform of a sharp top hat function of height $\widehat{\Delta \phi}(x^A)$,
\begin{align}\label{tophat}
    \Delta \tilde{\phi} \approx \widehat{\Delta \phi}(x^A) \frac{2}{\omega} \sin \left( \frac{\omega T}{2} \right), \qquad \omega \mathcal{T} \ll 1.
\end{align}
Using this approximation in \eqref{F1}, we find
\begin{align}
    I_1(T) &\approx C \int_{0}^{\omega_c} d\omega  \frac{4}{\omega} \sin^2(\omega T/2) \coth\left( \frac{\pi \omega}{\kappa} \right) \label{I1s1} \\
    & = C \int_{0}^{\omega_c T} d\lambda \frac{4}{\lambda} \sin^2(\lambda/2) \coth\left( \frac{\pi \lambda}{\kappa T} \right),\label{I1s2}
\end{align}
where
\begin{align}\label{C}
C =\frac{1}{\pi} \int \widehat{\Delta \phi}^2 dS
\end{align}
is the constant we call the decohering flux.  In passing from \eqref{I1s1} to \eqref{I1s2} we made the change of variables $\lambda =\omega T$.  

The analysis now bifurcates according to whether $\kappa$ vanishes or not.  For $\kappa \neq 0$ we may replace $\coth x$ by its small-$x$ behavior $1/x$ in the large-$T$ limit,
\begin{align}
    I_1(T) & \approx C \kappa T \frac{4}{\pi} \int_{0}^{\omega_c T} d\lambda  \frac{\sin^2(\lambda/2)}{\lambda^2} \\
    &\approx C \kappa T, \qquad T \to \infty, \ \ \kappa \neq 0,\label{F1nonextremal}
\end{align}
noting that $\omega_c T\to \infty$ with the intermediate scaling \eqref{scales} and using the integral $\int_0^\infty \sin^2(\lambda/2)/\lambda^2 d\lambda=\pi/4$.  

Alternatively, in the $\kappa \to 0$ limit we may replace $\coth x$ by its large-$x$ value of $1$,
\begin{align}
    I_1(T) & \approx  4 C \int_{0}^{\omega_c T} d\lambda  \frac{\sin^2(\lambda/2)}{\lambda} \\
    &\approx 2 C \log T, \qquad T\to \infty, \ \ \kappa = 0,\label{F1extremal}
\end{align}
where in the second step we drop constant terms, including $\log \omega_c$.  In both cases the arbitrary scale $\omega_c$ disappears from the leading behavior at large $T$. 

Since our calculation assumes a bifurcate Killing horizon, strictly speaking it applies only for non-zero $\kappa$.  The expression ``$\kappa=0$'' in \eqref{F1extremal} (and elsewhere below) is shorthand for the assumption $\kappa T \ll 1$, which establishes the hierarchy $\mathcal{T} \ll T \ll 1/\kappa$.  One might worry that the Aretakis instability \cite{Aretakis:2012ei,Casals:2016mel,Gajic:2023uwh} will make this hierarchy difficult to achieve, since growth occurs on timescales of order $\kappa^{-1}$ for near-extremal black holes \cite{Gralla:2016sxp}.  However, the field itself decays during this time, albeit at a slow, power-law rate. 

Now consider the second integral \eqref{F2}.  For analyzing this integral it is helpful to write $\Delta \phi$ as
\begin{align}\label{Deltaphimodel}
    \Delta \phi|_{\mathcal{H}^-_E}=g_1(u+T/2,x^A)-g_2(u-T/2,x^A),
\end{align}
where $g_1(u,x^A)$ and $g_2(u,x^A)$ are smooth, $T$-independent functions that transition from $0$ to $\widehat{\Delta \phi}(x^A)/2$ in a region of size $\Delta u \sim \mathcal{T}$ near $u=0$.  These functions describe the details of the transition periods, including (for example) oscillations from quasi-normal mode ringing.  The Fourier transform of \eqref{Deltaphimodel} is
\begin{align}
    \Delta \tilde{\phi}=\tilde{g}_1 e^{-i \omega T/2} -\tilde{g}_2  e^{i \omega T/2},
\end{align}
where $\tilde{g}_1(\omega,\theta,\phi)$ and $\tilde{g}_2(\omega,\theta,\phi)$ are the Fourier transforms of $g_1$ and $g_2$, respectively.  The magnitude squared is thus
\begin{align}\label{Deltaphisquaredmodel}
    |\Delta \tilde{\phi}|^2=|\tilde{g}_1|^2 + |\tilde{g}_2|^2 + \tilde{g}_1 \overline{\tilde{g}}_2 e^{i \omega T} + \tilde{g}_2 \overline{\tilde{g}}_1 e^{-i \omega T}.
\end{align}
In the second integral \eqref{F2} we have $\omega T \gg 1$ everywhere on the domain of integration and by \eqref{Deltaphisquaredmodel} the $T$-dependence of $|\Delta \tilde{\phi}|^2$ is only in the form of a rapid oscillation.  These oscillations integrate to zero, leaving
\begin{align}
    I_2 \approx \frac{1}{\pi} \int dS \int_{\omega_c}^\infty d\omega (|\tilde{g}_1|^2 + |\tilde{g}_2|^2) \omega \coth\left( \frac{\pi \omega}{\kappa} \right).
\end{align}
Although the integrand is now independent of $T$, the integral can still depend on $T$ through the lower-limit $\omega_c$, which has an intermediate scaling, such as $\omega_c \sim 1/\sqrt{T}$.  However, we have already seen that $\omega_c$ disappears from the leading, large-$T$ approximation for the other integral $I_1$ (linearly diverging in the non-extremal case, and logarithmically diverging in the extremal case).  Since the full integral $I_1+I_2$ is manifestly independent of $\omega_c$, it follows that $I_2$ is also independent of $\omega_c$ at these orders in the large-$T$ expansion.  But since $I_2$ can only have $T$-dependence through $\omega_c$, we conclude that $I_2$ makes no contribution at these orders in the large-$T$ expansion.

The leading large-$T$ behavior of $\langle N_e\rangle$ \eqref{split} thus comes entirely from the low-frequency integral $I_1$.  Collecting the results from Eqs.~\eqref{F1nonextremal} and \eqref{F1extremal}, we have
\begin{align}
    \langle N_e \rangle_{\mathcal{H}^-_E} \approx C \begin{cases} \kappa T & \kappa \neq 0 \\ 2\log T, & \kappa = 0 \end{cases},\label{Nefinal}
\end{align}
holding for $T \gg \mathcal{T}$.  The log of a dimensionful quantity appears because we have dropped subleading constant contributions; these depend on the details of the transitions.

Recall from \eqref{largetdecoherence} that the exponential of $-\langle N_e\rangle/2$ gives the final value of the of the inner product $|\!\braket{\psi_L|\psi_R}\!|$ of the left and right states of the experiment.  If there are no significant contributions to the entangling photon flux from other boundaries, then Eq.~\eqref{Nefinal} provides the decoherence rate.  We see that in the non-extremal case the dependence on the separation time $T$ is exponential,
\begin{align}\label{nonextremalresult}
     |\!\braket{\psi_L|\psi_R}\!| & = e^{-\frac{C}{2} \kappa T},
\end{align}
which we can interpret as saying that the state decoheres exponentially.  On the other hand, in the extremal case we have a slower, power-law decoherence
\begin{align}\label{extremalresult}
     |\!\braket{\psi_L|\psi_R}\!| & = (T/T_0)^{-C},
\end{align}
where we now include an undetermined constant $T_0$.  This constant is non-universal and depends on the details of the transition.  By contrast, the decohering flux $C$ can be determined from the stationary field equations via Eq.~\eqref{C}.  It remains to calculate the decohering flux in situations of physical interest.

\section{Observer on the symmetry axis of a rotating black hole}\label{sec:Kerr}

The Kerr metric in Boyer-Lindquist coordinates reads
\begin{align}\label{Kerr_Boyer_Lindquist}
    ds^2 &= - \left(1 - \frac{2Mr}{\Sigma}\right) dt^2 + \frac{\Sigma}{\Delta} dr^2 \\ & + \Sigma d\theta^2  -\frac{4M ar \sin^2\theta}{\Sigma}dtd\phi \nonumber \\ &+ \left(r^2 + a^2+ \frac{2 M a^2 r \sin^2\theta}{\Sigma} \right)\sin^2\theta d\phi^2
\end{align}
where $\Delta = r^2 + a^2 - 2 M r$ and $\Sigma = r^2 + a^2 \cos^2\theta$.  The horizon radius $r_+$, angular velocity $\Omega_H$, and surface gravity $\kappa$ are given by 
\begin{align}
    r_+ & = M + \sqrt{M^2-a^2} \\
    \Omega_H & = \frac{a}{r_+^2+a^2} \\
    \kappa & = \frac{\sqrt{M^2-a^2}}{2M r_+}.
\end{align}
The Boyer-Lindquist coordinates are not regular on the horizon.  To describe the past horizon $\mathcal{H}^-$ we introduce
\begin{align}
u & = t - r_*, \label{outgoing u}\\
\psi^- & = \phi - r^\# \label{outgoing psi},
\end{align}
where $dr_*/dr = (r^2+a^2)/\Delta$ and $d r^\#/dr = a/\Delta$.  See Eqs.~(5.62) and (5.63) in Ref.~\cite{poisson_relativists_2004} for explicit expressions for $r_*$ and $r^\#$.  The coordinates $(u,r,\theta,\psi^-)$ are regular on the past horizon, but they are not constant on its generators, which instead rotate with angular velocity $\Omega_H$.  We therefore introduce an additional angle
\begin{align}
    \bar{\phi} = \psi^- - \Omega_H u, \label{co-rotating_angle}
\end{align}
which remains constant on the generators of the past horizon.  We will refer to the set $(u,r,\theta,\bar{\phi})$ as ``horizon-adapted coordinates''.  These satisfy the construction described in Sec.~\ref{sec:bifurcate} with $x^A=(\theta,\bar{\phi})$.  

The induced metric on the horizon is
\begin{align}
    ds^{2}_{\mathcal{B}} & = q_{AB} dx^A dx^B \\ 
    & = \Sigma_+ d\theta^2 + \Sigma_+^{-1} (r_+^2+a^2)^2\sin^2\theta d\bar{\phi}^2,
\end{align}
where the $+$ indicates evaluation at $r=r_+$.  The area element is
\begin{align}\label{dSKerr}
    dS = (r_+^2 + a^2) \sin \theta d \theta d\bar{\phi}.
\end{align}

The horizon-generating Killing field takes the form 
\begin{align}
    \xi = \frac{\partial}{\partial t} + \Omega_H \frac{\partial}{\partial \phi} = \frac{\partial}{\partial u}.
\end{align}
The partial derivatives in the first equality refer to Boyer-Lindquist coordinates, whereas the partial derivative in the second equality refers to horizon-adapted coordinates. 

The state of a quantum field near a black hole formed from gravitational collapse is described at late times by the Unruh vacuum \cite{unruh_notes_1976}, which corresponds to choosing a set of modes that are positive frequency with respect to affine time $U$ when evaluated on the past horizon $\mathcal{H}^-$ (along with additional modes that vanish there).  Using this same set of modes in the framework of Sec.~\ref{sec:experiment} describes the superposition experiment near a black hole formed from gravitational collapse.  Sec.~\ref{sec:bifurcate} showed that under these circumstances the decoherence rate is determined by the integral of the square of the stationary difference field on the horizon, $\widehat{\Delta \phi}$, as in Eq.~\eqref{C}.  Our task is to compute $\widehat{\Delta \phi}$ in cases of physical interest.

We will consider point sources described by a charge $q(\tau)$ on a worldline $x^\mu(\tau)$ (where $\tau$ is proper time),
\begin{align}\label{KG source}
    \rho = \int q(\tau) \frac{\delta^{(4)}(x- x(\tau))}{\sqrt{-g}} d\tau.
\end{align}
Notice that unlike in the electromagnetic case, the charge can be time-dependent.  For a constant point charge $q_0$ held at radius $r_0$ on the symmetry axis of a Kerr black hole, we have
\begin{align}
    \rho = \frac{\alpha q_0}{2\pi \Sigma} \delta(r-r_0)\delta(1-\cos\theta),
\end{align}
where $\alpha=\sqrt{g_{tt}}$ is the ``redshift factor'', 
\begin{align}\label{lapse squared KG alice}
    \alpha^2 = \frac{r_0^2+a^2 -2 M r_0}{r_0^2+a^2}.
\end{align}
The stationary solution $\phi_c = \phi_c(r,\theta)$ satisfies
\begin{align}\label{KG_point_charge}
    &\frac{\partial}{\partial r}\left(\Delta \frac{\partial \phi_c}{\partial r} \right) + \frac{1}{\sin\theta}\frac{\partial}{\partial \theta}\left(\sin\theta\frac{\partial \phi_c}{\partial r}\right)  \nonumber\\= &-2 \alpha q_0 \delta(r-r_0)\delta(1-\cos\theta).
\end{align}
The solution has been given in integral form in Ref.~\cite{linet_stationary_1977} as Eq.~(10) with $s=0$.  (Our $\phi_c$ is identified with $- 2\alpha q_0 f$ in that reference.)  Evaluating the integral gives
\begin{align}\label{phicKG}
    \phi_c(t,r,\theta,\phi) &= \frac{\alpha q_0}{\mathcal{R}}, 
\end{align}
where $\mathcal{R}$ is defined as
\begin{align}
    \mathcal{R}^2 &= \left[(r_0 - M) - (r-M)\cos\theta\right]^2 + \Delta \sin^2\theta. \label{radial separation KG kerr}
\end{align}

Notice that the constant prefactor in the KG Coulomb solution \eqref{phicKG} is the redshifted charge $\alpha q_0$.  In particular, the charge inferred at infinity as the coefficient of the $1/r$ falloff is $\alpha q_0$, which is not equal to the locally defined charge $q_0$.  As an extreme example, when a charge $q_0$ is slowly lowered into the black hole, the field everywhere outside the black hole decreases in strength and ultimately vanishes when the charge enters the horizon.  There is no obstacle to this physical process since there is no conservation law for the scalar charge.  The disappearance of the field is a  manifestation of the ``no scalar hair'' property of black holes \cite{Bekenstein:1971hc}.  

These properties make the KG spatial superposition experiment behave differently from the EM case.  To adhere most closely to the DSW electromagnetic setup, we would consider a point charge $q_0$ held in superposition at two slightly different radii $r_L$ and $r_R$, which results in a dipolar difference source.  However, if the dipole is defined in the Lorentz-invariant way as the charge $q_0$ times the proper distance $d$, then the field actually falls off like $1/r$ at infinity, not as $1/r^2$ as it does in electromagnetism.  In essence, the KG dipole still behaves like a monopole.

We therefore find it more convenient to simply work with a monopole superposition, which is possible since Alice can manipulate the value of the KG charge as a function of time.  That is, we consider the conceptually simpler experiment in which the particle position is definite but the value of its charge is placed in superposition and later restored to the original value.  We will still use the labels ``left'' and ``right'', but the two semiclassical sources share a single position $r_0$, differing only in the time-evolutions $q_R(\tau)$ and $q_L(\tau)$ of their charges.  The difference source $\Delta q = q_R-q_L$ evolves from zero to some value $Q$ and back.  The horizon difference field  $\widehat{\Delta \phi}$ \eqref{widephidef} is given by the difference of Coulomb fields on the horizon,
\begin{align}
\widehat{\Delta \phi} = -\frac{\alpha Q}{r_0 - M - (r_+-M) \cos \theta}.
\end{align}
Using \eqref{dSKerr} we may compute the decohering flux as 
\begin{align}\label{CKerrKG}
    C = 4 Q^2 \frac{r_+^2+a^2}{r_0^2+a^2},
\end{align}
where we remind the reader that $Q$ is the difference in charge between the branches of the superposition.

In general there are also contributions to the decoherence from null infinity.  In  App.~\ref{sec: inner product} we show that the expected number of entangling particles on $\mathcal{I}^-$ at large $T$ is given by $\brkt{N_e}_{\mathcal{I}^-} = 2 C_\infty \log T$ \eqref{Neinfinity}, where $C_\infty$ is the analogous decohering flux on null infinity \eqref{Cinfinity}.  From Eq.~\eqref{phicKG} we compute the difference of Coulomb fields at $r \to \infty$ to be $
    \widehat{\Delta \phi}_\infty = -\alpha Q$, and the integral \eqref{Cinfinity} gives
\begin{align}\label{CinfinityKerr}
    C_\infty = 4 \alpha^2 Q^2.
\end{align}
This gives rise to a power-law contribution to large-$T$ decoherence going as $(T/T_0)^{-C_\infty}$, similar to the extremal horizon result \eqref{extremalresult}.

In the non-extremal case the decoherence from the horizon is $\exp[-(1/2)C\kappa T]$ by \eqref{nonextremalresult} and always dominates the power-law contribution from null infinity.  On the other hand, in the extremal case the horizon's contribution is also a power law, and the dominant effect is the one with the larger decohering flux.  Comparing \eqref{CKerrKG} and \eqref{CinfinityKerr}, we find (as expected) that the horizon dominates when the observer is sufficiently near the horizon, while infinity dominates when the observer is sufficiently far away.  Precise equality occurs at the critical value $r_0=M+\sqrt{M^2+r_+^2}$.

We began this paper by reviewing a heuristic argument involving Bob making measurements from behind a horizon, which suggests the presence of a decoherence effect associated with horizons.  The precise calculation shows there is also a decoherence associated with null infinity (in the KG case), and it is interesting to consider whether a similar heuristic applies.  While no single Bob can gather information at infinite distance, we can still imagine an army of Bobs stationed on a distant sphere in the distant future, i.e., a congruence of observers approaching null infinity.  The question of whether the Bobs gather finite information in the limit depends on falloff conditions, and we do not attempt to analyze it directly.  However, the non-zero decoherence result suggests that for the KG field, the Bobs can indeed gather which-path information from far away.  It would be very interesting to make this correspondence precise.

\section{Electromagnetic decoherence}\label{sec:EM}

The EM case is highly analogous to the KG case, except for some additional subtleties related to the choice of gauge.  We will first establish some some general properties of a horizon-adapted gauge, before proceeding to perform the calculation.

\subsection{Horizon-adapted gauge}\label{sec:gauge}

Following DSW we work in a gauge where $A_{\mu}$ vanishes when contracted with the horizon generator.  Since we work on the past horizon, the condition takes the form 
\begin{align}
    \xi^\mu A_\mu |_{\mathcal{H}^-} = 0.
\end{align}
In the horizon-adapted coordinates of Sec.~\ref{sec:bifurcate} we may write
\begin{align}\label{Gauge_U}
    A_U|_{\mathcal{H}^-} = 0
\end{align}
or
\begin{align}
    A_u|_{\mathcal{H}^-_E} = 0.
\end{align}
We will call this a horizon-adapted gauge.

In App.~\ref{sec:killing} we show that, in the horizon-adapted gauge, the pullback of Maxwell's equation to the horizon takes the form of a total derivative,
\begin{align}\label{Maxwell_total_derivative}
     \frac{\pd}{\pd u}\left(\frac{1}{2}\epsilon^{AB}{}^*\! F_{AB} - \nabla_{A} A^A \right) = 0.
\end{align}
Integrating then gives
\begin{align}\label{divA0}
    \nabla_A A^A = \frac{1}{2}  \epsilon^{AB} \ \! {}^*\!F_{AB}  + f(x^A),
\end{align}
in terms of an integration constant $f$.  Under the residual gauge freedom $A_A \to A_A + \nabla_A g$ with $g=g(x^A)$, we have
\begin{align}
    f \to f + q^{AB}\nabla_A \nabla_B g.
\end{align}
We can thus utilize the residual gauge freedom to make $f(x^A)$ lie complement of the image of the Laplacian.  By theorem 4.13a of Ref.~\cite{aubin_nonlinear_1998}, for compact manifolds (in our case, the horizon cross-section $\mathcal{C}$) this complement is just the set of constant functions.  We may thus set $f$ to be a constant,
\begin{align}\label{divA2}
    \nabla_A A^A = \frac{1}{2}  \epsilon^{AB} \ \! {}^*\!F_{AB} + f_0.
\end{align}
After integration, we see that $f_0$ is proportional to the charge enclosed in the compact horizon, 
\begin{align}\label{divA2}
    f_0 = \frac{1}{\mathcal{A}}\int_{\mathcal{C}} \frac{1}{2} \epsilon^{AB} {}^*\!F_{AB} \ \! dS = \frac{1}{\mathcal{A}}\int_{\mathcal{C}} {}^*F,
\end{align}
where $dS$ is the natural area element on $\mathcal{C}$ and $\mathcal{A}=\int_\mathcal{C} dS$ is the cross-sectional area.  If there is no charge in the compact horizon then we have simply 
\begin{align}\label{divA}
    \nabla_A A^A = \frac{1}{2}  \epsilon^{AB} \ \! {}^*\!F_{AB}.
\end{align}
For non-compact horizons with suitable falloff conditions, we expect that the complement of the image of the Laplacian will either be empty or can be removed with a similar physical assumption.  

For simplicity we will assume that $f_0=0$ so that \eqref{divA} holds.  However, it is trivial to repeat the analysis of the paper with $f_0$ included.  One simply assumes that the left and right branches of the superposition share the same $f_0$, which corresponds physically to the statement that the superposition was created from a single initial state.  (In the compact-horizon case, $f_0$ is interpreted as total charge.)  The constant $f_0$  then cancels out of the right-left difference field and hence does not appear in the final decoherence rate.  This cancellation occurs in our treatment of null infinity in App.~\ref{sec:EMnull}, where the charge cannot be assumed to vanish separately in the left and right branches.

A second equation analogous to \eqref{divA} can be derived much more easily.  From the definition of the field strength we have
\begin{align}
    \pd_A A_B - \pd_B A_A& = F_{AB}.\label{FAB}
\end{align}
Contracting this equation with $\epsilon^{AB}$ produces a magnetic analog of \eqref{divA}.  Displaying these two equations together gives a pleasing pair,
\begin{align}
    q^{AB}\nabla_A A_B & =  \frac{1}{2} \epsilon^{AB} \ \! {}^*\!F_{AB} \label{Ersource} \\
    \epsilon^{AB} \nabla_A A_B & = \frac{1}{2}\epsilon^{AB} F_{AB}. \label{Brsource}
\end{align}
The right-hand-sides provide invariant notions of the radial electric and magnetic fields. 

These equations suggest that both components of $A_A$ can be determined from the field strength $F_{\mu \nu}$ at each time $u$.  We can make this explicit by introducing electric and magnetic Hertz potentials $\mathcal{E}(u)$ and $\mathcal{B}(u)$,
\begin{align}\label{EBdecomp}
    A_A = \nabla_A \mathcal{E} + \epsilon_{AB}\nabla^B \mathcal{B}.
\end{align}
Each potential is then sourced by its corresponding radial field,
\begin{align}
    \nabla^2 \mathcal{E} & =  - \frac{1}{2} \epsilon^{AB} \ \! {}^*\!F_{AB} \label{Eeqn} \\
    \nabla^2\mathcal{B} & =  -\frac{1}{2}\epsilon^{AB} F_{AB}, \label{Beqn}
\end{align}
where $\nabla^2=q^{AB}\nabla_A \nabla_B$ is the Laplacian on the horizon cross-section.  We define the inverse Laplacian $\nabla^{-2}$ so that the integral of $\nabla^{-2} f$ vanishes for any $f$ in the domain.  (This is always possible by theorem~4.13e of Ref.~\cite{aubin_nonlinear_1998}.)  We then invert as
\begin{align}
    \mathcal{E}=-\frac{1}{2} \nabla^{-2} \left( \epsilon^{AB} \ \! {}^*\!F_{AB} \right) \label{Eformal} \\
    \mathcal{B}=-\frac{1}{2} \nabla^{-2} \left( \epsilon^{AB} F_{AB} \right). \label{Bformal}
\end{align}
We expect similar expressions to hold in the non-compact case provided that there are suitable falloff conditions.

\subsection{Entangling photons}\label{sec:entangling_photons}

With the above facts established, the KG analysis of Secs.~\ref{sec:QFT}--\ref{sec:Kerr} generalizes straightforwardly to EM fields. 
 The modes satisfy the free Maxwell equation,
\begin{align}
    \nabla_\mu F^{\mu \nu} = 0,
\end{align}
and are normalized according to gauge-invariant,\footnote{\label{gauge_confusion}Under $A_\mu \to A_\mu + \partial_\mu \Lambda$, Eq.~\eqref{EM inner product} changes by boundary terms, which are the integrals of $F_1 \Lambda$ and $F_2 \Lambda$.  We do not address these subtleties in the mode quantization argument, but the final results involve the inner product evaluated on solutions where such terms would vanish---see discussion below Eq.~\eqref{DeltaA}.} surface-independent product 
\begin{align}\label{EM inner product}
    (A_1,A_2) & = -\frac{i}{4\pi}\int_{\Sigma}d^3x \sqrt{h} n^\mu \left(\overline{F}_{1\mu \nu} A_2^\nu-F_{2\mu \nu}\overline{A}_1^{\nu}\right).
\end{align}
In the presence of a source $J^\mu$ we have 
\begin{align}
\nabla_\mu F^{\mu \nu} = -4\pi J^\nu.
\end{align}
We make the same assumptions of Sec.~\ref{sec:experiment} with sources $J_R^\mu$ and $J_L^\mu$ instead of $\rho_R$ and $\rho_L$, and use identical arguments to show the decoherence rate is given by $e^{-\langle N_e \rangle/2}$ \eqref{largetdecoherence} with
\begin{align}\label{Ne2EM}
    \langle N_e \rangle = (K \Delta A,K\Delta A),
\end{align}
analogously to \eqref{Ne2}.  Here $\Delta A$ is the right-left difference of the retarded-minus-advanced fields,
 \begin{align}\label{DeltaA}
    \Delta A & \equiv (A_R^{\rm ret} - A_R^{\rm adv}) - ( A_L^{\rm ret} - A_L^{\rm adv}),
\end{align}
where we suppress the tensor index and use notation analogous to \eqref{Deltaphi}.  
Note that the associated field strength $\Delta F$ is the retarded minus advanced solution arising from a source $J^\mu_R-J^\mu_L$ of compact spacetime support.  This ensures that $\Delta F$ has compact spatial support on any Cauchy surface, and therefore that the product \eqref{Ne2EM} is gauge-invariant.

The steps of Sec.~\ref{sec:bifurcate} also generalize.  On the horizon in the horizon-adapted gauge we have
\begin{align}\label{EM horizon inner product}
    (A_1, A_2&)_{\mathcal{H}^-} =  \frac{i}{4\pi}\int d S \int_{-\infty}^\infty dU  \\ 
    & q^{AB} \left(\ \!\overline{A}_{1A} \partial_U A_{2B} - A_{2 A} \partial_U \overline{A}_{1 B} \right),\nonumber
\end{align}
which closely parallels the corresponding KG expression \eqref{KGonH}.  The ensuing steps proceed identically until we establish the analog of \eqref{NeKilling},
\begin{align}\label{NeKilling_EM}
    \langle N_e \rangle_{\mathcal{H}^-_E} = \frac{1}{4 \pi^2} \int dS \int_{0}^\infty \frac{\omega d\omega}{2\pi} |\Delta \tilde{A}|^2  \coth\left( \frac{\pi \omega}{\kappa} \right),
\end{align}
where
\begin{align}
    |\Delta \tilde{A}|^2 = q^{AB} \Delta \overline{\tilde{A}}_A \Delta \tilde{A}_B|_{\mathcal{H}^-_E},
\end{align}
with $\tilde{A}_A$ is the Killing-time Fourier transform on the Horizon [analogous to \eqref{Deltaphitilde}],
\begin{align}
    \Delta \tilde{A}_A = \int_{-\infty}^\infty \Delta A_A|_{\mathcal{H}^-_E} e^{-i \omega u} du.
\end{align}

In the KG case we proceeded to estimate the integral based on the assumption that $\Delta \phi$ transitions from zero to a constant $\widehat{\Delta \phi}$ and back, with the transitions occurring over a timescale $\mathcal{T}$ and the constant period occurring for $T \gg \mathcal{T}$ (Fig.~\ref{fig:setup}).  In the EM case the same assumptions will be valid for the field strength $F_{\mu \nu}$, but the behavior in a given gauge $A_A$ need not mimic that of the field strength.  However, we have shown in Sec.~\ref{sec:gauge} that the horizon-adapted gauge can be constructed locally in time from the field strength (at least for compact horizons, and presumably in the non-compact case with suitable falloff), meaning that it will in fact share the properties described above.  We may thus perform the estimate identically, arriving at the exact same results \eqref{nonextremalresult} and \eqref{extremalresult}, 
\begin{align}
|\!\braket{\psi_L|\psi_R}\!| & = \begin{cases} e^{-\frac{C}{2} \kappa T}, & \kappa \neq 0 \\
    (T/T_0)^{-C} &\kappa=0 \end{cases},
\end{align}
where the decohering flux $C$ is now given by
\begin{align}\label{CEM}
    C = \frac{1}{4\pi^2} \int q^{AB} \widehat{\Delta A}_A \widehat{\Delta A}_B \ \! dS,
\end{align}
analogously to \eqref{C}.  The quantity $\widehat{\Delta A}$ is defined analogously to \eqref{widephidef},
\begin{align}\label{wideAdef}
    \widehat{\Delta A}(x^A) = -\left( A_R^{\rm stat} - A^{\rm stat}_L \right)|_{\mathcal{H}},
\end{align}
where (with tensor indices suppressed) $A_R^{\rm stat}$ and $A_L^{\rm stat}$ are stationary solutions in horizon-adapted gauge in the presence of stationary sources $J_R^{\rm stat}$ and $J_L^{\rm stat}$ (respectively) corresponding to the superposition.

To derive the formula for $C$ given in the introduction as Eq.~\eqref{CEMformal}, we use the solution \eqref{Eformal} and \eqref{Bformal} for the electric and magnetic potentials in the decomposition \eqref{EBdecomp} for the horizon vector potential.  Plugging this decomposition into \eqref{CEM}, we find that cross-terms are total derivatives.  These may be dropped for a horizon of compact support like that of the Kerr spacetime, and also for non-compact horizons in spacetimes with suitable decay properties (such as Rindler) given the assumption of a localized source outside the horizon.

\subsection{Calculation in Kerr}

The charge-current of a point source in a curved spacetime is given by
\begin{align}\label{EM source}
    J^\mu = \int e u^\mu \frac{\delta^{(4)}(x- x(\tau))}{\sqrt{-g}} d\tau.
\end{align}
For a stationary charge on the symmetry axis of a Kerr black hole, we have
\begin{align}
    J = \frac{e}{2\pi \Sigma} \delta(r-r_0)\delta(1-\cos\theta) \frac{\partial}{\partial u}.
\end{align}
The stationary, regular solution with charge $e$ can be called the Coulomb field $F_{\mu \nu}^c$.  A vector potential is given in closed form in Eq.~(9) of Ref.~\cite{B_Leaute_1982}.  However, this potential is poorly behaved on the horizon and its well-behaved components are not in the horizon-adapted gauge.  While we could in principle proceed by finding an explicit gauge transformation to a regular, horizon-adapted gauge, we find it more convenient to compute the gauge-invariant field strength and reconstruct a a suitable $A_A^c$ by solving  Eqs.~\eqref{Ersource} and \eqref{Brsource}.  In the axisymmetric case of present interest, these equations become
\begin{align}
    \frac{1}{\sqrt{q}}\partial_\theta\left(\sqrt{q}q^{\theta\theta} A^c_\theta \right) & =  -\frac{1}{2} \epsilon^{AB} \ \! {}^*\!F^c_{AB} \\
        \frac{1}{\sqrt{q}}\partial_\theta A^c_{\bar{\phi}}& = \frac{1}{2}\epsilon^{AB} F^c_{AB}, 
\end{align}
holding on the horizon $r=r_+$.  The integrals can be done in closed form, and after some calculation we find 
\begin{align}
    A_A^c|_{\mathcal{C}} &= \frac{-e (r_+ - M)\sin\theta}{(a^2+r_0^2)(r_0-M-(r_+-M) \cos \theta)}\nonumber\\&\times\bigg[\left(r_0 r_+ + a^2\cos\theta\right) \left(d\theta\right)_A \label{Ahor} \\ & +\frac{a (r_+^2 +a^2)(r_+- r_0 \cos\theta)\sin\theta}{r_+^2 + a^2\cos^2\theta} \left(d\bar{\phi}\right)_A \bigg], \nonumber
\end{align}
imposing regularity on the pole to obtain a unique solution.

Eq.~\eqref{Ahor} refers to the Coulomb field of a single point charge $e$.  For the superposition experiment, a charge $e$ is held in a spatial superposition of two nearby radii $r_0 \pm \epsilon/2$.  We require the difference of the two vector potentials, which by linearity is the vector potential due to the difference of the sources.  Since the proper separation is $\sqrt{g_{rr}} \epsilon$, the difference source is an effective dipole $p = e \sqrt{g_{rr}} \epsilon$.  In the limit $\epsilon \to 0$ (fixing $p$) the difference field is 
\begin{align}
    A^{\rm stat}_{R,\mu} - A^{\rm stat}_{L,\mu} = \frac{e}{\sqrt{g_{rr}}|_{r=r_0,\theta=0}} \left.\frac{\partial A_\mu^c}{\partial r_0}\right|_{e=1}.
\end{align}
Using Eq.~\eqref{Ahor} we can obtain $\widehat{\Delta A}$ \eqref{wideAdef} and evaluate the decohering flux integral \eqref{CEM}.  We find
\begin{widetext}
\begin{align}\label{CKerrEM}
   C &= p^2\frac{ (r_+^2+a^2)(r_0- r_+)(r_0r_+ - a^2)}{6\pi(r_0^2+a^2)^3r_+}\biggr[6\frac{r_+^2+a^2}{r_+^2-a^2}\log\left(\frac{r_+(r_0-r_+)}{r_0r_+ - a^2}\right)\nonumber\\ & + \frac{a^6+a^4 \left(-5 r_0^2+18 r_0r_+-14 r_+^2\right)+a^2 r_+ \left(6 r_0^3-26 r_0^2 r_++18 r_0 r_+^2+r_+^3\right)+r_0^2 r_+^3 (6 r_0-5 r_+)}{(r_0-r_+)^2 \left(a^2-r_0 r_+\right)^2}\biggr].
\end{align}
\end{widetext}
This function is plotted in Fig.~\ref{fig:C} as a function of black hole spin. 

We can obtain simpler expressions in a few different limits.  First we we consider the non-spinning case $a=0$,
\begin{align}\label{Cschw}
    C &= \frac{4 M^2 p^2}{3 \pi r_0^5}\bigg(2M\frac{3r_0 - 5M}{r_0 - 2M} \nonumber \\
    & \qquad \qquad + 3 (r_0 - 2M)\ln\big(1-\frac{2M}{r_0} \big)\bigg).
\end{align}
We can further expand at large $r_0$,
\begin{align}\label{Cfarnonspinning}
    C = \frac{32}{3 \pi} \frac{p^2 M^4}{r_0^6 } + O(r_0^{-7}).
\end{align}
This expression may be compared to analogous results in Ref.~\cite{Danielson:2022tdw}, which quotes a decoherence timescale in Eq.~(15).  In light of Eq.~\eqref{nonextremalresult} we define the decoherence timescale $T_D = 2/(C\kappa)$.  Using $\kappa=1/(4M)$ and Eq.~\eqref{Cfarnonspinning} for $C$, and noting that our $p$ equals their $qd$, we confirm the scaling of Eq.~(15) of Ref.~\cite{Danielson:2022tdw}.  Noting our units $\epsilon_0=1/(4\pi)$, we find that their Eq.~(15) should contain a prefactor $3\pi^2$ in order to correspond to our definition of $T_D$.  This is the precise version of their order of magnitude estimate.

Alternatively we can consider a near-horizon limit. Letting $b=2\sqrt{2M(r_0- 2M)} +O((r_0- 2M)^{3/2})$ denote the proper distance to the horizon, we find
\begin{align}
    C = \frac{2 p^2}{3 \pi b^2} + O(b^0).
\end{align}
This may be compared with analogous results for Rindler spacetime in Ref.~\cite{Danielson:2022sga}, noting that the acceleration $a$ of a near-horizon observer is related to proper distance $b$ by $a=1/b+O(b)$.  Since Ref.~\cite{Danielson:2022sga} quotes a rate in proper time, we also use $\kappa T = a\tau + O(b)$ to convert our decoherence rate $T_D$ to a proper decoherence rate $\tau_D=T_D(\kappa/a)=2/(C a)$.  This $\tau_D$ may be compared to $T_D$ in Eq.~(3.27) of Ref.~\cite{Danielson:2022sga}; we confirm the scaling and find that a prefactor of $12\pi^2$ is required to agree with the precise rate.

It is also instructive to look at the large $r_0$ limit at finite spin,
\begin{align}
    C = \frac{2 p^2 (r_+^2 - a^2)^2(r_+^2 + a^2)}{3 \pi r_+^2 r_0^6 } + O(r_0^{-7}).
\end{align}

Finally, we consider extremal limits.  Introducing $\epsilon=\sqrt{1-a^2/M^2}$ and letting $\epsilon \to 0$ at fixed $M$ and $r_0$, we find
\begin{align}
    C = \frac{8 p^2 M^4(M^2 - 2M r_0 + 2r_0^2)}{3\pi (M - r_0)^2 (M^2 + r_0^2)^3} \epsilon^2 + O(\epsilon^{3}).
\end{align}
The flux vanishes like $\epsilon^2$, which is linear in $a$, as seen in Fig.~\ref{fig:C}.  

Alternatively, we can take the extremal limit while simultaneously approaching the horizon. Defining $x_0=(r_0-r_+)/r_0$ and taking the extremal limit $\epsilon \to 0$ with $x_0 \sim \epsilon$ gives a finite limit,
\begin{align}
    C = p^2 \frac{ \epsilon^2}{3\pi M^2x_0( x_0 + 2 \epsilon)}.
\end{align}
If time is also rescaled like $1/\epsilon$, then this limit applied to the metric yields the the``nearNHEK'' patch \cite{Bredberg:2009pv} of the NHEK metric \cite{Bardeen:1999px}.  Here we have used the definitions of Ref.~\cite{gralla_particle_2015}.

In the case of KG fields there was also a contribution to decoherence from null infinity.  This occurs because the Coulomb fields are non-zero at infinity in the sense that the $1/r$ part of the difference field is non-zero.  This occurs for the monopole case we studied as well as the KG dipole case, as discussed in Sec.~\ref{sec:Kerr}.  However, in the EM case the dipole field decays like $1/r^2$ and there is no contribution to decoherence from null infinity.  This may be seen explicitly from the general expression \eqref{CEMformalinfinity} derived in App.~\ref{sec: inner product}.

\section*{Acknowledgements}
It is a pleasure to thank Paul Anderson, Daine Danielson, Gautam Satishchandran, and Bob Wald and for helpful discussions.  This work was supported by NSF grants PHY-1752809 and PHY-2309191 to the University of Arizona.

\appendix

\section{Symplectic product on the Kerr exterior and contributions from null infinity}\label{sec: inner product}

In this appendix, we consider the exterior part of the Kerr spacetime, foliated by Cauchy surfaces of constant Boyer-Lindquist $t$, denoted $\Sigma_t$.  As $t \to -\infty$, the symplectic product splits into separate contributions from the past horizon $H^-_E$ and from past null infinity $\mathcal{I}^-$, with contributions from past timelike infinity assumed to vanish due to falloff. 
 The horizon contributions were estimated at large $T$ in the main body; here we estimate the large-$T$ contributions from null infinity.  We consider the KG and EM cases separately.  We find that there is a logarithmic contribution in the KG case, whereas the EM case is finite as $T \to \infty$.

\subsection{Klein-Gordon product}\label{sec: inner product breaks apart}

As already presented in the main text as Eq.~\eqref{KGproduct}, the KG product of two source-free solutions $\phi_1$ and $\phi_2$ on a timeslice $\Sigma_t$ is defined as
\begin{align}\label{Generic_inner_prod}
    (\phi_1, \phi_2)_t = i\int_{\Sigma_t} d^3x \sqrt{h} n^\mu \left(\ \! \overline{\phi}_1 \nabla_\mu \phi_2 - \phi_2 \nabla_\mu \overline{\phi}_1 \right),
\end{align}
where $n^\mu$ is the future-directed unit normal to $\Sigma_t$, and $h$ is the determinant of the induced metric.  It follows directly from the source-free Klein-Gordon equation that the inner product is independent of the surface $\Sigma_t$.

We may approach the past horizon by letting $r \to r_+$  at fixed $\{u, \theta, \bar{\phi}\}$, where these coordinates were defined in Sec.~\ref{sec:Kerr}.  Since $t \to -\infty$ in this limit, it represents a portion of the limiting $\Sigma_t$.  In this limit, a direct computation gives
\begin{align}\label{limit_3_volume_element_horizon}
     d^3 x \sqrt{h} n^\mu \to dS du \left(\frac{\pd}{\pd u}\right)^\mu,
\end{align}
where $dS$ is given in Eq.~\eqref{dSKerr} and $\left(\frac{\pd}{\pd u}\right)^\mu$ is the coordinate basis vector in the horizon-adapted coordinate system $\{u, r, \theta, \bar{\phi}\}$ (also equal to the horizon-generating Killing field).  The contribution to the KG product in this limit is thus
\begin{align}
    (\phi_1, \phi_2)_{\mathcal{H}^-_E} &= i\int_{-\infty}^\infty du \int dS \left(\ \!\overline{\phi}_1\pd_u \phi_2 - \phi_2\pd_u\overline{\phi}_1 \right)_{\mathcal{H}_E^-},
\end{align}
where the partial derivative refers to horizon-adapted coordinates.

We may approach past null infinity by letting $r \to \infty$ at fixed $(v, \theta, \phi)$, where $v$ is the advanced time $v=t+r_*$.  Since $t \to -\infty$ in this limit, it also represents a portion of the limiting $\Sigma_t$. In this limit, a direct computation gives
\begin{align}\label{limit_3_volume_element_null_infinity}
    d^3 x \sqrt{h} n^\mu \to r^2 d\Omega dv \left(\frac{\partial}{\partial v}\right)^\mu,
\end{align}
where $d\Omega = \sin\theta d\theta d\phi$ is the 2-sphere element and $\left(\frac{\partial}{\partial v}\right)^\mu$ is the coordinate basis vector in the advanced coordinate system $\{v,r,\theta,\phi\}$.  The volume element diverges, but this will be canceled in the inner product by falloff of the fields,\footnote{This falloff is clear in the mode solutions to the KG equation \cite{teukolsky_1972}.}
\begin{align}\label{null infinity field KG appendix}
    \phi(v,r,y^A) = \phi|_{\mathcal{I}^-}(v,y^A)r^{-1} + o(r^{-1}).
\end{align}
(We adopt the notation that evaluation of a scalar field at $\mathcal{I}^-$ represents the $1/r$ part.)  The contribution to the KG product in this limit is thus
\begin{align}\label{KG product null infinity}
    (\phi_1, \phi_2)_{\mathcal{I}^-} &= i\int_{-\infty}^\infty \! dv \int \! d\Omega \left(\ \! \overline{\phi}_1\pd_v \phi_2 - \phi_2 \pd_v\overline{\phi}_1 \right)_{\mathcal{I}^-}.
\end{align}
where the partial derivative refers to advanced coordinates. 
 The fields in the integrand are understood as the $r^{-1}$ parts defined in Eq.~\eqref{null infinity field KG appendix}.

Finally we consider the pieces of the limiting $\Sigma_t$ that are not captured by the horizon or null infinity.  Colloquially, these are the ``contributions from past timelike infinity''.  Although we do not formalize a proof, it is clear that these pieces do not contribute to the limiting KG product.  Extensive numerical evidence (beginning with \cite{price_nonspherical_1972}) shows that solutions with compact sources decay polynomially in time at fixed spatial coordinate.  Rigorous decay results have also been obtained in hyperboloidal slicings (Ref.~\cite{Dafermos_Rodnianski_Shlapentokh-Rothman_2016}, Corollary 3.1).  The time-reverse of these results applies to decay of the advanced field at early times, such that the retarded-minus-advanced field we consider should decay as $t \to -\infty$ away from $\mathcal{H}^-$ and $\mathcal{I}^-$.  We therefore assert that there is no contribution from timelike infinity,
\begin{align}
    \lim_{t \to -\infty} (\phi_1, \phi_2)_t = (\phi_1, \phi_2)_{\mathcal{H}_E^-} + (\phi_1, \phi_2)_{\mathcal{I}^-}.
\end{align}

\subsection{Decoherence from Null Infinity}\label{infinity decoherence appendix}

We now consider the superposition experiment setup of the main body.  In Secs.~\ref{sec:bifurcate} and \ref{sec:Kerr} we evaluated the contribution to decoherence from the past horizon, first in a general setup and then for an observer on the symmetry axis of the Kerr spacetime.  We now repeat this analysis for the contribution from null infinity.  The steps are highly analogous.

For the Unruh state, the non-vanishing mode functions are positive frequency with respect to $v$,
\begin{align}\label{KG in mode}
    \left.\phi^{\text{in}}_{\omega L}\right|_{\mathcal{I}^-} = \frac{e^{-i\omega v}}{\sqrt{4\pi\omega}}Y_{L}(y^A),
\end{align}
where $y^A = \{\theta, \phi\}$ and we remind the reader that evaluation at $\mathcal{I}^-$ is understood as taking the $1/r$ part, as in Eq.~\eqref{null infinity field KG appendix}.

The functions $Y_{L}(y^A)$ are assumed to be orthonormal on the (celestial) sphere,
\begin{align}
    \int  \overline{Y}_{L}(y^A)Y_{L'}(y^A)d\Omega = \delta_{L L'}.
\end{align}
In practice, the most useful angular mode functions depend on $\omega$ as well,\footnote{The scalar wave equation is separable in the Kerr background, with $\theta$ eigenfunctions equal to oblate spheroidal harmonics $S^m{}_{\ell}(-a^2\omega^2, \cos\theta)$ (See Eq.~(8) in Ref.~\cite{teukolsky_1972}).
Thus we have
\begin{align}
    Y_{\omega \ell m}(\theta, \phi) = \frac{e^{-i m \phi}}{\sqrt{4\pi}}S^m{}_{\ell}(-a^2\omega^2\!\!,\cos\theta).
\end{align}
} but our results are insensitive to the specific choice. According to the falloff \eqref{null infinity field KG appendix}, a field on $\mathcal{I}^-$ may be decomposed as
\begin{align}
    \left.\phi\right|_{\mathcal{I}^-}(v,x^A)=\int_{-\infty}^\infty d\omega\sum_L c_{\omega L}\left.\phi^{\text{in}}_{\omega L}\right|_{\mathcal{I}^-}(v,x^A)
\end{align}
for some mode coefficients $c_{\omega L}$. Using the same method as in Sec.~\ref{sec:KG product in affine time FT}, we find that
\begin{align}
    (K\phi&,K\phi)_{\mathcal{I}^-}  = \int_0^{\infty} d \omega \sum_L |c_{\omega L}|^2
     \\
    & = 2 \int d\Omega \int_0^\infty \frac{\omega d\omega}{2\pi} |\tilde{\Phi}(\omega,x^A)|^2\label{KG_number_fourier_transform_null_infinity}\\
    & = \! \frac{-1}{\pi} \! \! \int \! \! d\Omega d v_1 dv_2 \frac{\phi(v_1,x^A) \phi(v_2,x^A)}{(v_1-v_2 - i \epsilon)^2},\label{KG_number_double_smear_null_infinity}
\end{align}
where $\tilde{\Phi}(\omega,x^A)$ is now the Fourier transform of $\Phi$ on past null infinity $\mathcal{I}^-$, 
\begin{align}\label{DeltaphitildeScri}
    \tilde{\Phi} = \int_{-\infty}^\infty  \phi|_{\mathcal{I}^-} e^{-i \omega v} dv.
\end{align}

In the superposition experiment, the contribution to the entangling particles from null infinity is 
\begin{align}
    \brkt{N_e}_{\mathcal{I}^-} = (K\Delta \phi,K\Delta \phi)_{\mathcal{I}^-},
\end{align}
where $\Delta \phi$ is the left-right difference of the retarded-minus-advanced soluution, Eq.~\eqref{Deltaphi}.  The retarded solutions cancel at early times, so we have
\begin{align}
    \left.\Delta\phi\right|_{\mathcal{I}^-} =  -\left.\left(\phi_R^\text{adv}- \phi_L^\text{adv}\right)\right|_{\mathcal{I}^-}.
\end{align}
Assuming that Alice is a static observer,\footnote{In a more general asymptotically flat setting (beyond the Kerr metric), we would require that she is on a Killing orbit of $\partial_v$, the Killing field generating $\mathcal{I}^-$.} we may estimate $(K\Delta \phi,K\Delta\phi)$
using the method of Sec.~\ref{sec:killing-time-estimate} applied to Eq.~\eqref{KG_number_fourier_transform_null_infinity}.  Since the advanced time $v$ agrees with Killing time, there is no factor of $\coth\left(\frac{\pi \omega}{\kappa}\right)$ [as in Eq.~\eqref{NeKilling}] appearing in Eq.~\eqref{KG_number_fourier_transform_null_infinity}. Therefore, the method of estimation in Sec.~\ref{sec:killing-time-estimate} applies with $\kappa = 0$ [See Eq.~\eqref{Nefinal}]. We obtain
\begin{align}\label{Neinfinity}
    \brkt{N_e}_{\mathcal{I}^-} = 2 C_\infty \log T
\end{align}
with 
\begin{align}\label{Cinfinity}
    C_\infty = \frac{1}{\pi}\int |\widehat{\Delta \phi}_\infty|^2 d\Omega,
\end{align}
where $\widehat{\Delta \phi}_\infty$ is the difference of the stationary Coulomb solutions,
\begin{align}\label{widephi_null_infinity}
    \widehat{\Delta \phi}_\infty(x^A) = -\left( \phi_R^{\rm stat} - \phi^{\rm stat}_L \right)|_{r\to \infty}.
\end{align}
We refer to $C_\infty$ as the decohering flux on null infinity.

\subsection{Electromagnetic decoherence at null infinity}\label{sec:EMnull}

We now consider electromagnetic decoherence in the Kerr spacetime.  The main differences in the EM case will be related to the choice of gauge.  As already given in the main text as Eq.~\eqref{EM inner product}, the EM symplectic product is
\begin{align}\label{EM inner product2}
    (A_1,A_2) & = -\frac{i}{4\pi}\int_{\Sigma_t}d^3x \sqrt{h} n^\mu \left(\overline{F}_{1\mu \nu}A_2^\nu-F_{2\mu \nu}\overline{A}_1^{\nu}\right).
\end{align}
Since the product is gauge-invariant (up to boundary terms---see footnote \ref{gauge_confusion}), we are free to use different gauges on different portions of $\Sigma_t$.  For the portion that limits to the past horizon as $t \to -\infty$, we use the horizon-adapted gauge of the main text.  For the portion that limits to null infinity we will use an analogous gauge, as follows.

We again work in advanced coordinates $(v, r, y^A)$ in the Kerr spacetime, letting $r\to\infty$ to reach $\mathcal{I}^-$. (However, our analysis does not use details of the Kerr metric in an essential way and extends naturally to more general asymptotically flat spacetimes.)  We denote the spherical metric by $\gamma_{AB}$, where in $\{\theta,\phi\}$ coordinates,
\begin{align}
    \gamma_{AB} = (d\theta)_A(d\theta)_B + \sin^2\theta (d\phi)_A (d\phi)_B.
\end{align}
The covariant derivative with respect to the 2-sphere is denoted as $\mathcal{D}_A$ and the 2-dimensional area element is $\epsilon_{AB}$.  We use orientation $\epsilon_{v r \theta \phi}=\sqrt{-g}$ and $\epsilon_{\theta \phi} = \sqrt{\gamma} =  \sin\theta$.

We assume that the electromagnetic field is smooth at null infinity in the sense of Eq.~(27) in Ref.~\cite{Satishchandran_2019},\footnote{We can convert formulas at future null infinity in \cite{Satishchandran_2019} to past null infinity by sending $u \to -v$.}
\begin{align}
    A_v &= O(r^0) \\ 
    A_r & = O(r^{-2}) \label{Arscri}\\
    A_A & = O(r^0).\label{AAscri}
\end{align}
In analogy with the horizon adapted gauge where $A_u=0$ on $\mathcal{H}^-_E$, we want $A_v$ to vanish near $\mathcal{I}^-$. We use the notation that a superscript $(n)$ indicates the $r^{-n}$ part of a quantity expanded at large $r$.  
 If $A_v$ is expressed as $A_v^{(0)} + A_v^{(1)}r^{-1} + O(r^{-2})$, we make a gauge transformation $A_\mu \to A_\mu + \partial_\mu \Lambda$ with $\Lambda = \int dv A_v^{(0)} +  \int dv A_v^{(1)} r^{-1} + O(r^{-2})$.  This transformation makes $A_v$ order $r^{-2}$, while preserving the falloff conditions \eqref{Arscri} and \eqref{AAscri} of $A_r$ and $A_A$,
\begin{align}    
A_v &= O(r^{-2}), \label{Av ansatz scri}\\ A_r &= O(r^{-2}), \label{Ar ansatz scri}\\ A_A &= A_A^{(0)}+O(r^{-1}).\label{AA ansatz scri}\end{align}
The field strength has falloff
\begin{align}
    F_{vr} &= F_{vr}^{(2)} r^{-2} + O(r^{-3})\\
    F_{vA} &= F_{vA}^{(0)} + O(r^{-1})\\
    F_{rA} &= O(r^{-2})\\
    F_{AB} & = F_{AB}^{(0)} + O(r^{-1})
\end{align}
with
\begin{align}
    F_{v A}^{(0)}& = \partial_v A_A^{(0)},\label{FvA relation AA null infinity} \\ F_{AB}^{(0)} &= \partial_A A_B^{(0)} - \partial_B A_A^{(0)}.\label{FAA relation AA null infinity}
\end{align}
It is convenient to introduce radial electric and magnetic fields on $\mathcal{I}^-$ as
\begin{align}
    E^{(2)}_r &\equiv \lim_{r\to \infty}\frac{1}{2}\epsilon^{AB} {}^*F_{AB} = -F_{vr}^{(2)},\label{radial electric field null infinity}\\
    B^{(2)}_r &\equiv \lim_{r\to \infty}\frac{1}{2}\epsilon^{AB} F_{AB} = -{}^*\!F^{(2)}_{vr},
    \label{radial magnetic field null infinity}
\end{align}
where we remind the reader that ${}^* F_{\mu \nu} = \frac{1}{2}\epsilon_{\mu \nu \rho \sigma} F^{\rho\sigma}$ is the Hodge dual of $F_{\mu \nu}$.

Assuming sufficient falloff towards timelike infinity, the symplectic product again splits into contributions from the past horizon and from past null infinity,
\begin{align}
    \lim_{t \to -\infty} (A_1, A_2)_t = (A_1, A_2)_{\mathcal{H}_E^-} + (A_1, A_2)_{\mathcal{I}^-}.
\end{align}
The horizon contribution was studied in the main body, and we now study the contribution from infinity.  A direct calculation shows that as $r \to \infty$ at fixed $v$, in our gauge we have
\begin{align}
    \sqrt{h}n^\mu F_{1\mu \nu}A_2^\nu \to - \sqrt{\gamma}\gamma^{AB}A_{2A}^{(0)}\partial_v A_{1B}^{(0)},
\end{align}
where $\gamma = \text{det}\gamma_{AB}$.  The limiting symplectic product is thus 
\begin{align}\label{EM product null infinity}
    &(A_1, A_2)_{\mathcal{I}^-} =\nonumber\\& \!\!\!\frac{i}{4\pi}\int_{-\infty}^\infty dv \int d\Omega \gamma^{AB} \left( \ \!\overline{A_{1A}^{(0)}}\partial_v A_{2B}^{(0)}- A_{2A}^{(0)}\partial_v \overline{A_{1B}^{(0)}} \ \!\right),
\end{align}
where $d\Omega$ is the volume element on the celestial sphere $d\Omega = \sqrt{\gamma} d^2 x^A$. It is convenient to work in this gauge because the symplectic product \eqref{EM product null infinity} is analogous to KG case \eqref{KG product null infinity}.  

We now repeat the analysis in Sec.~\ref{sec:gauge} for $\mathcal{I}^-$ instead of $\mathcal{H}^-_E$. The leading [$O(r^{-2})$] part of Maxwell's equations $\left(\frac{\pd}{\pd v}\right)^\nu\nabla_\mu F^{\mu}{}_{\nu} = 0$ gives
\begin{align}
    \frac{\pd}{\pd v}F_{vr}^{(2)} + \mathcal{D}^AF_{vA}^{(0)} = 0.
\end{align}
Using Eq.~\eqref{FvA relation AA null infinity} this becomes a total $v$-derivative, and we find
\begin{align}
    \mathcal{D}^AA_A^{(0)} = -F_{vr}^{(2)} + f(x^A)
\end{align}
in terms of an integration constant $f(x^A)$, analogous to Eq.~\eqref{divA0}.  We can change $f$ by the Laplacian of a scalar $g(x^A)$, i.e. $f \to f+\mathcal{D}^2 g$, using the residual gauge freedom $A^{(0)}_A \to A^{(0)}_A + \mathcal{D}_A g$.\footnote{These gauge transformations are sometimes called `large gauge transformations' \cite{Kapec:2015ena}.}  We can use this freedom to set $f$ equal to a constant $-e$,
\begin{align}\label{Electric part of AA null infinity}
    \mathcal{D}^AA_{A}^{(0)} = -F_{vr}^{(2)} - e.
\end{align}
Integrating over the sphere shows that $e$ is the total electric charge, 
\begin{align}
    e 
    = \lim_{r\to \infty}\frac{1}{4\pi}\int_{S^2} {}^*F,
\end{align}
When treating the horizon case we assumed for simplicity that the enclosed charge vanished, since a non-zero charge would cancel out when the right-left difference field is considered anyway.  Here we cannot take the charge to vanish, and we will see the right-left cancellation explicitly in Eq.~\eqref{CEMformalinfinity} below.

From Eqs.~\eqref{Electric part of AA null infinity} and \eqref{FAA relation AA null infinity} in the notation of Eqs.~ \eqref{radial electric field null infinity} and~\eqref{radial magnetic field null infinity}, we have
\begin{align}
    \gamma^{AB} \mathcal{D}_A A_B^{(0)} & = E_r^{(2)} - e  \label{construct A from Er}\\
    \epsilon^{AB} \mathcal{D}_A A_B^{(0)} & = B_r^{(2)},\label{construct A from Br}
\end{align}
Noting the definitions \eqref{radial electric field null infinity} and \eqref{radial magnetic field null infinity} of $E_r^{(2)}$ and $B_r^{(2)}$, these equations are analogous to Eqs.~\eqref{Ersource} and \eqref{Brsource}. A term analogous to $e$ does not appear in the horizon version because the charge enclosed in the horizon was assumed to vanish.

Eqs.~\eqref{construct A from Er} and \eqref{construct A from Br} show that  $A_A^{(0)}$ can be determined from the gauge-invariant field strength locally in time (at each $v$) along $\mathcal{I}^-$.  To see this explicitly, we introduce the decomposition
\begin{align}
    A^{(0)}_A = q_{AB} \mathcal{D}^B \mathcal{E} + \epsilon_{AB} \mathcal{D}^B \mathcal{B},\label{EBdecomp}
\end{align}
so that Eqs.~\eqref{construct A from Er} and \eqref{construct A from Br} become
\begin{align}
    \mathcal{D}^2 \mathcal{E} &= E_r^{(2)} - e ,\label{electric part null infinity}\\
    \mathcal{D}^2 \mathcal{B} & = B_r^{(2)}\label{magnetic part null infinity}.
\end{align}
Defining the inverse Laplacian $\mathcal{D}^{-2}$ so that the integral $\mathcal{D}^{-2}f$ vanishes for any $f$ in the domain, we have the explicit solution
\begin{align}
        \mathcal{E} &= \mathcal{D}^{-2}\left( E_r^{(2)} - e\right) \label{gimmeE} \\
    \mathcal{B} & = \mathcal{D}^{-2}B_r^{(2)}, \label{gimmeB}
\end{align}
from which the vector potential $A^{(0)}_A$ in the correct gauge can be constructed via Eq.~\eqref{EBdecomp}.  Notice that the $\ell=0$ part of $E^{(2)}_r$ (i.e., the charge $e$) is naturally removed so that the resulting function is in the domain of $\mathcal{D}^{-2}$.

With these results established, we may now repeat the steps of Sec.~\ref{infinity decoherence appendix} above.  The analog of Eq.~\eqref{KG_number_fourier_transform_null_infinity} is found to be
\begin{align}
    (KA,KA)_{\mathcal{I}^-} = \frac{1}{4 \pi^2} \int d\Omega \int_{0}^\infty \frac{\omega d\omega}{2\pi} \gamma^{AB} \tilde{A}_A^{(0)}\overline{\tilde{A}_B^{(0)}},
\end{align}
where again tilde denotes Fourier transform,
\begin{align}\label{DeltaphitildeScri}
    \tilde{A}_A = \int_{-\infty}^\infty A_A^{(0)} e^{-i \omega v} dv.
\end{align}
The expected number of entangling particles is given by plugging in the left-right difference of retarded minus advanced fields $\Delta A$ \eqref{DeltaA},
\begin{align}
    \brkt{N_e}_{\mathcal{I}^-} = (K\Delta A,K\Delta A)_{\mathcal{I}^-}.
\end{align}
Estimating at large $T$ then gives
\begin{align}
    \brkt{N_e}_{\mathcal{I}^-} &= 2 C_\infty \log T,
\end{align}
with
\begin{align}
    C_\infty &= \frac{1}{4\pi^2}\int \gamma^{AB}\widehat{\Delta A}_{\infty A}\widehat{\Delta A}_{\infty B} d\Omega,
\end{align}
where as usual $\widehat{\Delta A}_A$ is the difference of the stationary ``Coulomb fields'' associated with the stationary portions of the left and right branches of the superposition,
\begin{align}\label{widephi_null_infinity_EM}
    \widehat{\Delta A}_\infty(x^A) = -\left( A_R^{\rm stat} - A^{\rm stat}_L \right)|_{r \to \infty},
\end{align}
with tensor index $A$ suppressed.

If we introduce the difference radial fields $\widehat{\Delta E}_r$ and $\widehat{\Delta B}_r$ constructed from $\Delta A_{\infty}$ via 
Eqs.~\eqref{radial electric field null infinity} and \eqref{radial magnetic field null infinity} (dropping the subscript $\infty$ and superscript (2) for conciseness), noting Eqs.~\eqref{EBdecomp}, \eqref{gimmeE}, and \eqref{gimmeB}  we can write the decohering flux in a manifestly gauge-invariant manner as 
\begin{align}\label{CEMformalinfinity}
    C_\infty = \frac{1}{4\pi^2} \! \! \int \! \left[ (\nabla_A \mathcal{D}^{-2} \widehat{\Delta E_r})^2 + (\nabla_A \mathcal{D}^{-2} \widehat{\Delta B_r})^2 \right] \! dS,
\end{align}
where a total derivative cross-term vanishes, analogously to the horizon calculation discussed in the last paragraph of Sec.~\ref{sec:entangling_photons}.  Eq.~\eqref{CEMformalinfinity}
is precisely analogous to the formula \eqref{CEMformal} for horizon decoherence presented in the introduction.  Notice that the charge 
 $e$ has canceled out of the final expression, since $\widehat{\Delta E}_r$ has no net charge.

\section{Maxwell's equations on a Killing horizon}\label{sec:killing}

In this appendix we study Maxwell's equations on a bifurcate Killing horizon and show that the horizon component becomes a total derivative, Eq.~\eqref{Maxwell_total_derivative}.

\subsection{Killing horizon}

Following~\cite{Kay:1988mu}, we assume there exists a one-parameter isometry whose fixed points form an orientable, spacelike, co-dimension 2 (hence 2-dimensional) surface $\mathcal{B}$.  Let $\ell^\mu$ and $n^\mu$ be two (continuously chosen, future directed) null normals $\mathcal{B}$ satisfying $\ell \cdot n = -1$.  (For brevity we denote inner products by $v_\mu w^\mu=v \cdot w$.)  We extend $n^\mu$ off of $\mathcal{B}$ via the affinely-parameterized null geodesic equation,
\begin{align}\label{affine_n}
 n^\mu \nabla_\mu n^\nu = 0.
\end{align}
The null surface generated by $n^\mu$ is called the past horizon $\mathcal{H}^-$.  The portion of $\mathcal{H}^-$ in the causal past of $\mathcal{B}$ is the ``exterior'' past horizon $\mathcal{H}_E^-$, while the part in the future is the ``interior'' $\mathcal{H}_I^-$.  An analogous construction can be made for the future horizon $\mathcal{H}^+$, but we will not use it in this appendix.  In what follows, all equations are understood to be evaluated on $\mathcal{H}^-_E$.

Let $\xi^\mu$ denote the generator (Killing field) of the isometry.  As explained in Sec.~2 in Ref.~\cite{Kay:1988mu}, $\xi^\mu$ is tangent to $\mathcal{H}^{\pm}$.  On $\mathcal{H}^-_E$ it is related by a positive function $f$,
\begin{align}\label{killing_affine_vector_relation}
    n^\mu = f \xi^\mu.
\end{align}
We introduce time coordinates $u$ and $U$ on $\mathcal{H}^-_E$ by
\begin{align}
    n^\mu \nabla_\mu U &= 1&&\text{``affine time"}\\
    \xi^\mu \nabla_\mu u &= 1 && \text{``Killing time"}.
\end{align}
We choose $U$ such that the bifurcation surface $\mathcal{B}$ is at $U = 0$, while $U = -1$ coincides with $u = 0$. It is shown in Sec.~2 in Ref.~\cite{Kay:1988mu} that
\begin{align}
    U = -\exp(-\kappa u),
\end{align}
where $\kappa$ is the ``surface gravity''. The constancy of $\kappa$ on horizon can be seen using the same proof in Sec. 12.5 of Ref.~\cite{Wald:1984rg}. It follows immediately that 
\begin{align}\label{f_function_past_horizon}
 f =   \frac{1}{\kappa |U|} = \frac{1}{\kappa} \exp(-\kappa u).
\end{align}
The null geodesic congruence $n^\mu$ is also normal to the horizon. It follows that the geodesic congruence is expansion-less, shear free, and twist-less, i.e.,
\begin{align}\label{no_expansion_shear_free}
    v_1^\mu v_2^\nu\nabla_\mu n_\nu = 0
\end{align}
for all $v_1^\mu, v_2^\mu$ tangent to the horizon $\mathcal{H}_E^-$. This result again can be directly seen from an anologous argument leading to Eq.~(12.5.20) in Ref.~\cite{Wald:1984rg}.

If $x^A$ is a coordinate system on the bifurcation surface $\mathcal{B}$, we obtain a coordinate system $\{U, x^A\}$ (hence $\{u,x^A\}$) on $\mathcal{H}_E^{-}$ by letting $x^A$ be constant on the null geodesic generators.  If we adjoin a fourth cooordinate $r$ that is constant on $\mathcal{H}_E^-$, then on $\mathcal{H}_E^-$ we have
\begin{align}\label{generator}
    n^\mu = \left(\frac{\pd}{\pd U}\right)^\mu, && \xi^\mu = \left(\frac{\pd}{\pd u}\right)^\mu.
\end{align}
Since $\xi^\mu$ is a Killing field, we have $\partial_u g_{AB}=0$ and hence 
\begin{align}\label{gABconst}
    \partial_U g_{AB}=0.
\end{align}

\subsection{Null tetrad on $\mathcal{H}^-_E$}

We may obtain a null tetrad on $\mathcal{B}$ by selecting a normalized complex-null vector $m$ ($m \cdot m = \overline{m} \cdot \overline{m} = 0, m \cdot \overline{m} = 1$) that is orthogonal to $\ell$ and $n$.  We extend $m^\mu$ to all of $\mathcal{H}_E^-$ by Lie transport along the null generators,
\begin{align}\label{lUnU_commute}
    \mathcal{L}_n m^\mu = \left[ n, m\right]^\mu =  0.
\end{align}
This is equivalent to the statement that in the preferred coordinates $(U,r,x^A)$, the components $m^\mu$ are constant on horizon generators,
\begin{align}
    \frac{\pd}{\pd U}m^\mu=0.
\end{align}
In particular, $m$ is tangent to horizon cross-sections; its only non-zero components are $m^A$.  

The inner products $m\cdot n$, $m \cdot m$ and $m\cdot \overline{m}$ are preserved by the Lie transport.  For example,
\begin{align}\label{Lie_transport_metric}
    \mathcal{L}_n\left( m^\mu n^\nu g_{\mu \nu}\right) &= m^\mu n^\nu\mathcal{L}_n g_{\mu \nu} \nonumber\\
     &=  m^\mu n^\nu \left(\frac{\nabla_\mu f}{f} n_\nu +\frac{\nabla_\nu f}{f} n_\mu \right)\nonumber\\
     & = 0.
\end{align}
The second equality follows from Killing's equations $\nabla_\mu \xi_\nu + \nabla_\nu \xi_\mu = 0$ together with Eq.~\eqref{killing_affine_vector_relation}.\footnote{Eq.~\eqref{killing_affine_vector_relation} holds only on the horizon, but in Eq.~\eqref{Lie_transport_metric} the indices are contracted with vectors $m$ and $n$ which are tangent to the horizon.} The third equality holds because $m^\mu n_\mu = 0$ and  $m^\mu m_\mu = 0$ on $\mathcal{H}_E^-$.

Thus the conditions $m \cdot \overline{n} =1$,  $\overline{m} \cdot \overline{m}=0$, and $m \cdot \overline{m} =1$ hold everywhere on $\mathcal{H}^-_E$.  To form a complete tetrad basis on the horizon, we choose $\ell^\mu$ to be the future-directed null vector that is uniquely specified by $\ell \cdot n = -1$ and $m\cdot \ell = 0$. This defines a complex null tetrad on the entire $\mathcal{H}_E^-$.

For future use we list the standard relations obeyed by a null tetrad.  As shown above, the vectors are orthonormal,
\begin{align}
    \ell \cdot n &= -1 && m \cdot\overline{m} = 1 \\ 
    \ell \cdot \ell  &= 0 &&n \cdot n =0\\
    \ell \cdot m &= 0 && n\cdot m = 0 && m \cdot m = 0.
\end{align}
Since the dot products of tetrad members are constant, for any two tetrad members $X, Y \in \{\ell, n, m ,\overline{m}\}$, we have
\begin{align}\label{derivative_contract_tetrad}
    X^\mu \nabla_\nu Y_\mu = -Y^\mu \nabla_\nu X_\mu.
\end{align}
This is equivalent to the anti-symmetry of the Ricci rotation coefficients.  In particular, $X_\mu \nabla_\nu X^\mu = 0$. The metric is given by
\begin{align}\label{metric_intermsof_tetrad}
    g_{\mu \nu} = -\ell_\mu n_\nu -\ell_\nu n_\mu + m_\mu \overline{m}_\nu +m_\nu \overline{m}_\mu.
\end{align}
The volume elements for the 4-space $\epsilon$ and the cross-section $\epsilon^{(2)}$ are given by
\begin{align}
    \epsilon_{\mu \nu \rho \sigma} &= 24 i \ell_{\left[\mu\right.} n_{\nu}m_\rho \overline{m}_{\left.\sigma\right]} \label{4-volume}\\
    \epsilon^{(2)}_{\mu \nu} &= 2 i m_{\left[\mu\right.} \overline{m}_{\left.\nu\right]}. \label{2-volume}
\end{align}

We label the tetrad components of tensors with a lowered index; for example, $\ell^\mu n_\nu T_\mu{}^\nu = T_{\ell n}$.  All tetrad components are seen as lower indices; no raising and lowering of tetrad indices appears in this work.  Derivatives are written similarly; for example, $\nabla_\ell=\ell^\mu \nabla_\mu$.  When a derivative acts on a tensor with tetrad indices, the tetrad contraction is taken first.  For example, for  a tensor $T^{\mu_1\mu_2}{}_{\nu_1 \nu_2}$, we might write
\begin{align}\label{Hongji special}
    \nabla_m T^{\mu_1\mu_2}{}_{\ell n} = m^\lambda \nabla_\lambda\left(T^{\mu_1\mu_2}{}_{\nu_1 \nu_2}\ell^{\nu_1} n^{\nu_2}\right).
\end{align}

\subsection{Cross-sections $\mathcal{C}$}

The spatial two-surfaces obtained by Lie transport of the bifurcation surface will be called cross-sections $\mathcal{C}$. 
 The induced metric on a cross-section may be written
\begin{align}\label{2-metric}
    q_{\mu \nu} = m_\mu \overline{m}_\nu +\overline{m}_\mu m_\nu 
\end{align}
and satisfies
\begin{align}\label{2-complete}
    q_{\mu \lambda} q^{\lambda}{}_\nu = q_{\mu\nu}.
\end{align}
Since $q_{AB}=g_{AB}$, by \eqref{gABconst} the components are constant on horizon generators,
\begin{align}\label{qABconst}
    \frac{\pd}{\pd U} q_{AB} = 0.
\end{align}
It follows from orthonormality of $m$ that $q^{AB} = m^{A}\overline{m}^{B} + \overline{m}^{A}m^{B}$ is equal to the matrix inverse of $q_{AB}$. The derivative compatible with $q_{AB}$ is denoted $\nabla^{(2)}_A$.  For any one-form $v_\mu$, the components $v_A$ satisfy
\begin{align}\label{2fun}
    \nabla^{(2)}_A v_B = q_A{}^{\mu} q_B{}^{\nu}\nabla_\mu v_\nu.
\end{align}

\subsection{Maxwell's equation on $\mathcal{H}^-_E$}

We now consider the component of Maxwell's equations tangent to the horizon generators,
\begin{align}
n_\nu\nabla_\mu F^{\mu \nu} = 0.
\end{align}
Using the Leibniz rule we have
\begin{align}\label{Maxwell_of_interest}
      \nabla_\mu F^\mu{}_n - F^{\mu \nu} \nabla_\mu n_\nu = 0
\end{align}
in the notation of \eqref{Hongji special}. The two terms in Eq.~\eqref{Maxwell_of_interest} can be simplified as follows:
\begin{align}
     \nabla_\mu F^\mu{}_n & = F_{m n}\nabla_\mu \overline{m}^\mu
      + \nabla_{\overline{m}}F_{m n} + \text{c.c.} \nonumber \\ & \qquad - \nabla_n F_{\ell n} \label{first_term_maxwell}\\
     F^{\mu \nu}\nabla_\mu n_\nu &= F_{mn}\big(n^\mu \nabla_{\overline{m}}\ell_\mu + \overline{m}^\mu \nabla_{\ell}n_\mu\big) + \text{c.c.},\label{second_term_maxwell}
\end{align}
where $\text{c.c.}$ denotes the complex conjugate of the preceeding terms on that line.  We have used the antisymmetry of $F_{\mu \nu}$, the vanishing of the projection of $\nabla_\mu n_\nu$ onto the cross-section $\mathcal{C}$ [Eq.~\eqref{no_expansion_shear_free}] (in particular, $m^\mu m^\nu \nabla_\mu n_\nu = m^\mu \overline{m}^\nu\nabla_\mu n_\nu = 0$), antisymmetry of the Ricci rotation coefficients [Eq.~\eqref{derivative_contract_tetrad}], and the geodesic equation [Eq.~\eqref{affine_n}].

We will use two identities. The first identity reads
\begin{align}\label{Fnm_coeff_simplify}
    &\nabla_\mu \overline{m}^\mu -\left( n^\mu \nabla_{\overline{m}}\ell_\mu + \overline{m}^\mu \nabla_{\ell}n_\mu\right) = m^\mu \nabla_{\overline{m}} m_\mu,
\end{align}
where we have used Eq.~\eqref{lUnU_commute}, Eq.~\eqref{derivative_contract_tetrad}, and Eq.~\eqref{metric_intermsof_tetrad}. The second identity is
\begin{align}\label{2-divergence_in_terms_of_tetrad}
   \!\!\!\!\!\!\nabla^{(2)A} F_{n A} &=  q^{\mu \nu} \nabla_\mu F_{n \nu}\\
   &=\nabla_{\overline{m}} F_{n m} + F_{n m} m^\mu \nabla_{\overline{m}} \overline{m}_\mu  + \text{c.c.}, \nonumber
\end{align}
where we used Eqs.~\eqref{lUnU_commute}, \eqref{derivative_contract_tetrad}, \eqref{2-metric} and Eq.~\eqref{2fun}.

Substituting Eqs.~\eqref{first_term_maxwell} and \eqref{second_term_maxwell} into \eqref{Maxwell_of_interest}, using  \eqref{Fnm_coeff_simplify} and \eqref{2-divergence_in_terms_of_tetrad} we find
\begin{align}\label{Final_Maxwel}
    -\nabla_n F_{\ell n} + \nabla^{(2)A} F_{n A} = 0.
\end{align}

We now work in coordinates $(U,r,x^A)$ using the  horizon-adapted gauge $A_U=0$.  Noting $n=\pd_U$ \eqref{generator}, we have 
\begin{align}\label{FnA appendix}
    F_{nA} = \partial_U A_A.
\end{align}
Since the spatial metric $q_{AB}$ is independent of $U$ \eqref{qABconst}, the partial derivative $\pd_U$ commutes with the spatial covariant derivative $\nabla^{(2)}$ in \eqref{Final_Maxwel}.  We therefore find
\begin{align}\label{Maxwell_total_derivative}
     \frac{\pd}{\pd U}\left( F_{\ell n} - \nabla^{(2)A} A_A\right) =0.
\end{align}
Using Eqs.~ \eqref{4-volume} and \eqref{2-volume}, it is easy to check that the tetrad component $F_{\ell n}$ can be expressed as
\begin{align}\label{Er_def}
    F_{\ell n} = \frac{1}{2}\epsilon^{(2)\mu \nu}\ \!{}^*\!F_{\mu \nu} = \frac{1}{2}\epsilon^{(2)AB}\ \!{}^*\!F_{AB},
\end{align}
using  ${}^* F_{\mu \nu} = \frac{1}{2}\epsilon_{\mu \nu \rho \sigma} F^{\mu \nu}$.  Integrating Eq.~\eqref{Maxwell_total_derivative} and using \eqref{Er_def} gives us \eqref{divA0} in the main body.

\bibliographystyle{utphys}
\bibliography{References.bib}
\end{document}